\documentclass[11pt]{article}

\usepackage{acl}
\usepackage{times}
\usepackage{latexsym}
\usepackage[T1]{fontenc}
\usepackage[utf8]{inputenc}
\usepackage{microtype}
\usepackage{graphicx}
\usepackage{amsmath}
\usepackage{amssymb}
\usepackage{booktabs}
\usepackage{pifont}
\usepackage{multirow}
\usepackage[table,dvipsnames]{xcolor}
\usepackage{subcaption}
\usepackage{algorithm}
\usepackage{algorithmic}

\usepackage{enumitem}
\usepackage{listings}
\usepackage{tcolorbox}
\tcbuselibrary{skins,breakable}

\tcbset{
    promptbox/.style={
        enhanced,
        breakable,
        colback=blue!3,
        colframe=blue!40!black,
        fonttitle=\bfseries,
        coltitle=black,
        colbacktitle=blue!15,
        attach boxed title to top left={yshift=-2mm, xshift=5mm},
        boxed title style={sharp corners, size=small},
        sharp corners,
        boxrule=0.5pt,
        left=6pt,
        right=6pt,
        top=6pt,
        bottom=6pt,
    },
    vtoprompt/.style={
        promptbox,
        colback=green!3,
        colframe=green!40!black,
        colbacktitle=green!15,
    },
    charmprompt/.style={
        promptbox,
        colback=orange!3,
        colframe=orange!40!black,
        colbacktitle=orange!10,
    },
    redialprompt/.style={
        promptbox,
        colback=purple!3,
        colframe=purple!40!black,
        colbacktitle=purple!10,
    },
    starprompt/.style={
        promptbox,
        colback=teal!3,
        colframe=teal!40!black,
        colbacktitle=teal!15,
    },
    museprompt/.style={
        promptbox,
        colback=cyan!3,
        colframe=cyan!40!black,
        colbacktitle=cyan!10,
    },
    inspiredprompt/.style={
        promptbox,
        colback=pink!5,
        colframe=pink!40!black,
        colbacktitle=pink!10,
    },
}

\newcommand{\method}{\textsc{Harpo}}
\newcommand{\starmod}{\textsc{Star}}
\newcommand{\charm}{\textsc{Charm}}
\newcommand{\bridge}{\textsc{Bridge}}
\newcommand{\maven}{\textsc{Maven}}
\newcommand{\vto}{\textsc{Vto}}
\newcommand{\val}[2]{#1$\pm$#2}
\newcommand{\best}[2]{\textbf{#1$\pm$#2}}
\newcommand{\gain}[1]{\textcolor{ForestGreen}{#1\%}}

\title{HARPO: Hierarchical Agentic Reasoning for User-Aligned Conversational Recommendation}

\author{
Subham Raj$^{1}$ \quad Aman Vaibhav Jha$^{1}$ \quad Mayank Anand$^{2}$ \quad Sriparna Saha$^{1}$  \\
$^{1}$Department of Computer Science and Engineering, Indian Institute of Technology Patna, India \\
$^{2}$Indian Institute of Information Technology Allahabad, India \\
\texttt{\{subham\_2221cs25, 2201ai54\_aman, sriparna\}@iitp.ac.in} \\
\texttt{anandmayank698@gmail.com}
}

\begin{document}
\maketitle

\begin{abstract}
Conversational recommender systems (CRSs) operate under incremental preference revelation, requiring recommendation decisions under uncertainty. While recent LLM-based approaches achieve strong performance on proxy metrics such as Recall@K and BLEU, they often fail to deliver high-quality, user-aligned recommendations in practice, as they optimize intermediate objectives like retrieval accuracy or fluent generation rather than recommendation quality itself. We propose HARPO (Hierarchical Agentic Reasoning with Preference Optimization), an agentic framework that reframes conversational recommendation as a structured decision-making process optimized for multi-dimensional recommendation quality. HARPO integrates (i) hierarchical preference learning that decomposes recommendation quality into interpretable dimensions (relevance, diversity, satisfaction, and engagement) with context-dependent weighting; (ii) deliberative tree-search reasoning guided by a learned value network evaluating candidate paths on predicted quality; and (iii) domain-agnostic reasoning abstractions through Virtual Tool Operations and multi-agent refinement. We evaluate HARPO on ReDial, INSPIRED, and MUSE, demonstrating consistent improvements over strong baselines on recommendation-centric metrics while maintaining competitive response quality. Our code is available at \href{https://harpo-bench.github.io}{https://harpo-bench.github.io}.
\end{abstract}

\section{Introduction}
Conversational recommender systems (CRSs) aim to assist users in discovering items that match their preferences through natural language interaction. Unlike traditional recommendation systems that rely on static user profiles or historical behavior, CRSs operate in a sequential setting where user preferences are revealed incrementally through dialogue. As a result, these systems must interpret nuanced and often underspecified user intent and make recommendation decisions under uncertainty.

\begin{figure}
    \centering
    \includegraphics[width=1.7\linewidth]{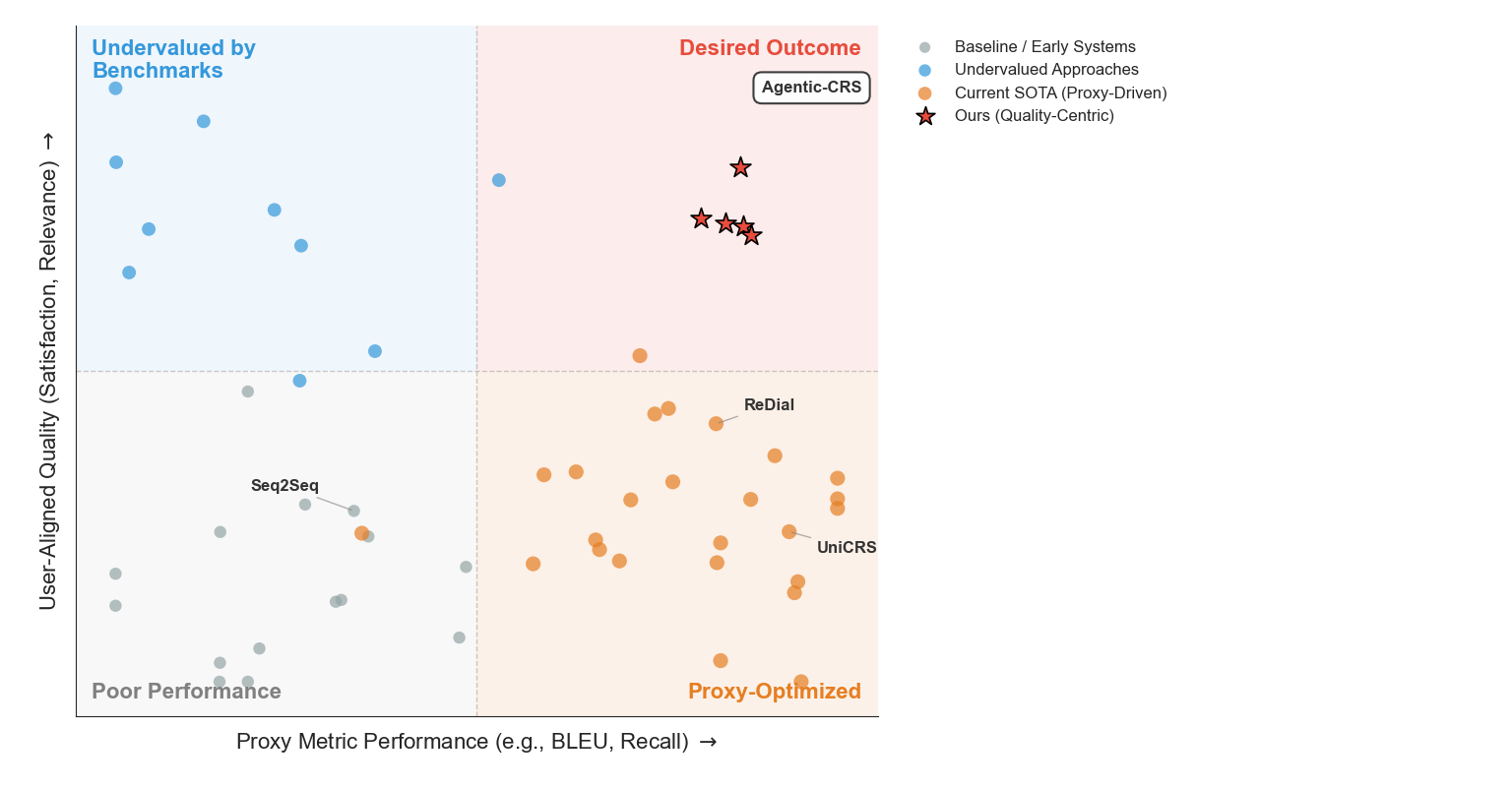}
    \caption{Conceptual illustration of the evaluation landscape for conversational recommender systems. The horizontal axis denotes performance on proxy metrics (e.g., BLEU, Recall), while the vertical axis represents user-aligned recommendation quality (e.g., satisfaction and relevance). Existing systems often achieve high proxy scores with limited alignment to user preferences. Positions are illustrative rather than measured values.}
    \label{fig:placeholder}
\end{figure}

Recent advances in large language models have significantly improved conversational recommendation performance~\cite{wang2022towards,dao2024broadening}. Modern CRS approaches, including retrieval-augmented and generation-based systems~\cite{li2018towards, hayati2020inspired, wang2022towards}, achieve strong results on widely used benchmarks, with high scores on metrics such as Recall@K, BLEU, and tool invocation accuracy. As illustrated in Figure~\ref{fig:placeholder}, these results create an impression that conversational recommendation is largely solved, despite limited alignment with user-centric recommendation quality.

However, strong proxy-metric performance does not necessarily translate into high-quality recommendations from a user's perspective. A system may retrieve appropriate items and generate fluent responses, yet still fail to capture nuanced user intent. For instance, a request for ``something casual for a summer wedding'' may be interpreted as everyday casual wear rather than context-appropriate wedding attire. While such responses score well on automatic metrics, they often lead to low user satisfaction in practice.

\begin{table*}[t]
\centering
\small
\setlength{\tabcolsep}{6pt}
\begin{tabular}{lccccc}
\toprule
\rowcolor{gray!12}
\textbf{Method} 
& \textbf{Recall@K} 
& \textbf{BLEU} 
& \textbf{Tool Acc.} 
& \textbf{Explicit Quality Modeling} 
& \textbf{User-Aligned Objective} \\

\midrule

ReDial-style CRS 
& \cellcolor{green!12}\checkmark 
& \cellcolor{green!12}\checkmark 
& \cellcolor{gray!10}-- 
& \cellcolor{red!12}$\times$ 
& \cellcolor{red!12}$\times$ \\

UniCRS           
& \cellcolor{green!12}\checkmark 
& \cellcolor{green!12}\checkmark 
& \cellcolor{green!12}\checkmark 
& \cellcolor{red!12}$\times$ 
& \cellcolor{red!12}$\times$ \\

DCRS             
& \cellcolor{green!12}\checkmark 
& \cellcolor{green!12}\checkmark 
& \cellcolor{green!12}\checkmark 
& \cellcolor{red!12}$\times$ 
& \cellcolor{red!12}$\times$ \\

\textbf{HARPO (ours)}     
& \cellcolor{green!18}\checkmark 
& \cellcolor{green!18}\checkmark 
& \cellcolor{green!18}\checkmark 
& \cellcolor{green!18}\checkmark 
& \cellcolor{green!18}\checkmark \\

\bottomrule
\end{tabular}
\caption{Comparison of evaluation objectives across representative conversational recommender systems. Existing CRS methods are primarily evaluated using proxy metrics, whereas HARPO explicitly optimizes user-aligned recommendation quality.}
\label{tab:eval-objectives}
\end{table*}

This gap reveals a fundamental misalignment between how conversational recommender systems are trained and evaluated and the actual objective of conversational recommendation. Existing approaches~\cite{li2018towards, hayati2020inspired, wang2022towards} primarily optimize proxy objectives such as lexical overlap or retrieval of ground-truth items, which only weakly correlate with user-aligned recommendation quality~\cite{jannach2021survey, gao2021advances}. Table~\ref{tab:eval-objectives} illustrates this reliance on proxy objectives, with explicit modeling of recommendation quality largely absent.

To address this limitation, conversational recommendation is framed as a structured decision-making problem~\cite{wei2022chain, yao2023tree} rather than a byproduct of response generation or tool execution. From this perspective, a system should explicitly reason over multiple candidate recommendation strategies, evaluate their expected quality, and select recommendations based on user-aligned criteria rather than proxy signals alone.

HARPO is proposed as an agentic framework that operationalizes this perspective by combining deliberative reasoning with explicit preference optimization across multiple dimensions of recommendation quality, including relevance, diversity, predicted user satisfaction, and engagement. Through structured reasoning and learned quality evaluation, HARPO enables systems to explore, compare, and refine recommendation decisions before generating final responses. It is evaluated on multiple conversational recommendation benchmarks, including ReDial~\cite{li2018towards}, INSPIRED~\cite{hayati2020inspired}, and MUSE~\cite{wang2025muse}.

In summary, HARPO makes the following contributions:
\begin{itemize}
    \item \textit{A fundamental misalignment in conversational recommender systems is identified, showing that proxy metrics fail to reflect user-aligned recommendation quality.}
    
    \item \textit{HARPO is introduced as an agentic framework that formulates conversational recommendation as a structured decision-making problem.}
    
    \item \textit{Multi-dimensional modeling of recommendation quality is incorporated to directly optimize user-aligned objectives such as relevance, diversity, and satisfaction.}
    
    \item \textit{A quality-centric evaluation perspective is proposed and empirically validated across multiple benchmarks.}
\end{itemize}

The remainder of the paper is organized as follows. Section~2 reviews related work. Section~3 details the HARPO framework. Section~4 reports experimental results. Section~5 concludes the paper.

\section{Related Work}
We review prior work in conversational recommendation, reasoning in language models, preference optimization, and tool-augmented generation, emphasizing the reliance on proxy objectives rather than explicit recommendation quality.

\subsection{Conversational Recommender Systems}

Conversational recommender systems (CRSs) have evolved from early constraint-based dialogue systems to neural architectures supporting open-ended interaction. Early approaches framed recommendation as a structured decision process, relying on attribute-based queries and explicit constraint elicitation to narrow candidate sets~\cite{christakopoulou2016towards, sun2018conversational}. ReDial~\cite{li2018towards} established the modern generation-based CRS paradigm, enabling natural language interaction while reinforcing reliance on automatic proxy metrics from recommendation and dialogue modeling.

Subsequent work incorporated external knowledge to improve recommendation accuracy and mitigate cold-start issues~\cite{chen2019towards, zhou2020improving}. Pre-trained language models further unified conversational recommendation architectures, enabling joint dialogue understanding, reasoning, and recommendation~\cite{wang2022barcor, wang2022towards}, with more recent methods integrating retrieval augmentation and multimodal grounding~\cite{dao2024broadening, wei2025mscrs}. Beyond conversational settings, multimodal recommendation has been explored through multitask learning frameworks that jointly optimize genre classification and rating prediction~\cite{raj2023multi}, learn unified user-movie representations from visual and textual modalities~\cite{raj2025multimodal}, and incorporate genre-aware scoring for domain-specific settings~\cite{mondal2023genre}; however, these approaches operate on static user profiles rather than the incremental preference revelation central to CRS. Despite broader advances in fluency, retrieval accuracy, and tool usage, existing CRSs are still primarily trained and evaluated using proxy metrics such as Recall@K and BLEU, which reward technical correctness but do not explicitly capture alignment with nuanced user intent or subjective satisfaction, motivating the need for quality-centric evaluation and optimization frameworks.

\subsection{Reasoning and Preference Optimization}
Explicit reasoning mechanisms such as chain-of-thought prompting and tree-based search improve performance on multi-step decision problems~\cite{wei2022chain, yao2023tree}. 
In parallel, preference optimization methods align model outputs with human judgments through learned reward signals~\cite{ouyang2022training, rafailov2023direct}. 
However, existing approaches typically optimize correctness-based or single-scalar objectives, leaving open how reasoning can be guided toward multi-dimensional, user-aligned outcomes. 
HARPO bridges this gap by combining deliberative reasoning with hierarchical preference learning, enabling value-guided exploration over candidate recommendation strategies based explicitly on predicted recommendation quality.

\subsection{Tool-Augmented Language Models}
Augmenting language models with external tools has proven effective for tasks requiring access to up-to-date information, specialized computation, or structured data sources~\cite{schick2023toolformer,qin2023toolllm}. In conversational recommendation, tools facilitate interaction with product catalogs, user histories, and knowledge graphs that cannot be fully captured within model parameters.

However, most existing tool-augmented approaches rely on domain-specific tools and training, limiting transferability and complicating evaluation. This tight coupling between tools and tasks further reinforces proxy-metric-driven benchmarks, as evaluation becomes dependent on specific tool implementations rather than underlying recommendation behavior.

To mitigate this issue, Virtual Tool Operations (VTOs) are introduced to decouple high-level reasoning from domain-specific tools. By defining domain-agnostic operations that are mapped to concrete tools at runtime, VTOs support transferable reasoning and more consistent evaluation across domains. This abstraction mirrors interface-based design in software engineering, enabling models to reason about recommendation decisions independently of underlying tool details.

\section{The \method{} Framework}

\begin{figure*}
    \centering
    \includegraphics[width=\linewidth]{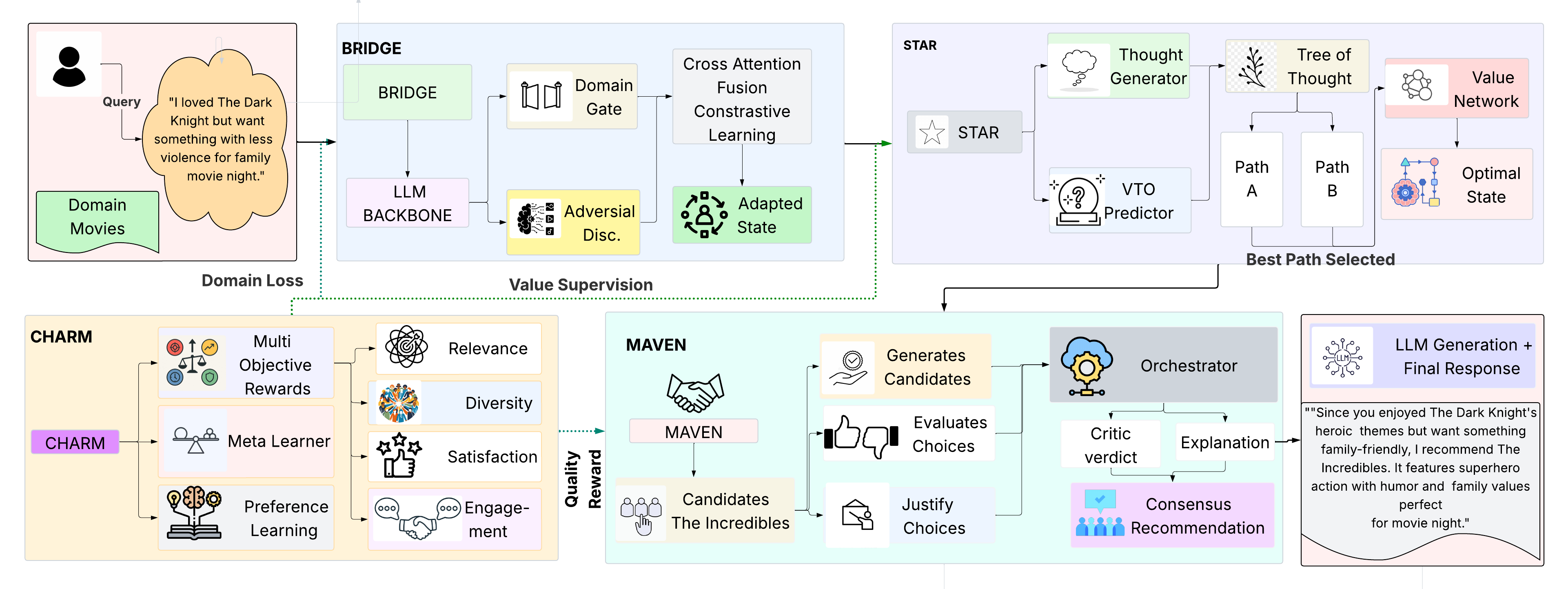}
    \caption{Overall architecture of the \method{} framework. The model integrates four components: \starmod{} for structured agentic reasoning, \charm{} for hierarchical preference optimization, \bridge{} for cross-domain transfer, and \maven{} for multi-agent refinement, all built on a shared language model backbone.}
    \label{fig:harpo-arch}
\end{figure*}

\method{} integrates four components within a unified architecture built upon a pre-trained language model backbone. We first formalize the optimization objective, then describe each component in detail, explaining the intuition and mathematical formulation of key mechanisms. An overview of the framework and the interaction between its components is shown in Figure~\ref{fig:harpo-arch}.
Additionally, the complete Virtual Tool Operations (\vto{}s) taxonomy, operation definitions, illustrative examples, and annotation procedure are provided in Appendix~\ref{app:vto}.

\subsection{Problem Formulation}

Let $\mathcal{C} = \{(u_1, r_1), \ldots, (u_{t-1}, r_{t-1}), u_t\}$ denote a conversation context consisting of user-system turn pairs and the current user utterance $u_t$. The system must generate a response $r_t$ containing recommendations. Let $\mathcal{V}$ denote the \vto{}s space and $\mathcal{D}$ the set of recommendation domains.

We seek a policy $\pi_\theta(r_t, \mathbf{v}_t \mid \mathcal{C}, d)$ that jointly generates responses and \vto{}s sequences, maximizing expected recommendation quality:
\begin{equation}
\theta^* = \arg\max_\theta \mathbb{E}_{\mathcal{C}, d} \left[ \mathcal{Q}(r_t, \mathcal{C}) \mid \pi_\theta \right]
\label{eq:objective}
\end{equation}
where $\mathbf{v}_t \in \mathcal{V}^*$ is the predicted \vto{}s sequence, $d \in \mathcal{D}$ is the domain, and $\mathcal{Q}(\cdot)$ measures recommendation quality.

The key departure from prior work lies in explicitly optimizing for $\mathcal{Q}$ rather than proxy objectives. Previous approaches optimize for item retrieval accuracy (maximizing probability of ground-truth items) or response quality (maximizing likelihood under reference responses), treating recommendation quality as an emergent property. Our formulation directly targets $\mathcal{Q}$, which we decompose into interpretable dimensions through \charm{}.

\subsection{\starmod{}: Structured Tree-of-Thought Agentic Reasoning}

STAR is a structured tree-of-thought\cite{yao2023tree} reasoning module that enables explicit exploration of multiple candidate recommendation strategies, guided by a learned value function estimating recommendation quality. Reasoning proceeds by expanding and evaluating candidate paths using beam search over structured reasoning states.

\paragraph{Reasoning State.}
Each node in the reasoning tree is represented as
\vspace{-1em}
\begin{equation}
s = (\mathbf{h}, \tau, \mathbf{v}, d),
\end{equation}

where $\mathbf{h}$ encodes accumulated dialogue and reasoning context, $\tau$ is the current natural language thought, $\mathbf{v}$ denotes predicted \vto{}s, and $d$ is the search depth. The hidden state $\mathbf{h}$ is computed from the full conversation context and the reasoning path, ensuring that value estimates reflect the complete trajectory.

\paragraph{Value Network.}
STAR employs a value network that predicts recommendation quality rather than task completion. Quality is decomposed into four dimensions—relevance, diversity, satisfaction, and engagement—each estimated by a dedicated head:
\begin{equation}
V_k(\mathbf{h}) = \sigma\!\left(\mathbf{W}_k^{(2)} \cdot \text{GELU}\!\left(\mathbf{W}_k^{(1)} \cdot \mathbf{h}\right)\right),
\end{equation}
where $\mathbf{W}_k^{(1)} \in \mathbb{R}^{d_h \times d_h}$ and $\mathbf{W}_k^{(2)} \in \mathbb{R}^{1 \times d_h}$ are learnable projection matrices, with the overall value computed as:
\begin{equation}
V(\mathbf{h}) = \sum_k \alpha_k V_k(\mathbf{h}),
\end{equation}

where $\alpha_k = \text{softmax}(\mathbf{w})_k$ are learnable weights over quality dimensions.

\paragraph{Thought Generation and Search.}
At each node, the language model generates $b$ candidate next steps,
\begin{equation}
\{(\mathbf{h}_i, \mathbf{v}_i, q_i)\}_{i=1}^{b} = \text{ThoughtGen}(\mathbf{h}),
\end{equation}
where $q_i$ is the predicted quality score for candidate $i$, which are filtered and explored using beam search with width $w$ and depth $D$:
\begin{equation}
s^* = \arg\max_{s \in \text{Beam}(s_0, w, D)} V(\mathbf{h}_s),
\end{equation}
where $s_0 = (\mathbf{h}_0, \tau_0, \mathbf{v}_0, 0)$ is the initial reasoning state corresponding to the conversation context prior to any reasoning step.
The final response is generated from the highest-valued reasoning path.

By training the value function with quality signals from \charm{}, STAR prioritizes reasoning paths that yield contextually appropriate, diverse, and satisfying recommendations, rather than optimizing task completion alone.

\subsection{\charm{}: Contrastive Hierarchical Alignment with Reward Marginalization}

CHARM optimizes recommendation quality by decomposing it into four interpretable dimensions—relevance, diversity, satisfaction, and engagement—and learning context-dependent weighting over these dimensions.

\paragraph{Hierarchical Reward Decomposition.}
Each quality dimension is modeled by a dedicated reward head:

\begin{equation}
R_k(\mathbf{h}) = \tanh\!\left(\mathbf{W}_k^{(2)} \cdot \text{GELU}\!\left(\mathbf{W}_k^{(1)} \cdot \mathbf{h}\right)\right),
\end{equation}
where $\mathbf{W}_k^{(1)} \in \mathbb{R}^{d_h \times d_h}$ and $\mathbf{W}_k^{(2)} \in \mathbb{R}^{1 \times d_h}$ are learnable projection matrices, and outputs are bounded to $[-1,1]$ for stable optimization. The total reward is computed as:

\begin{equation}
R(\mathbf{h}) = \sum_k w_k \cdot R_k(\mathbf{h}).
\end{equation}

\paragraph{Adaptive Weighting.}
Context-dependent weights are learned via meta-learning:

\begin{equation}
\mathbf{w} = \text{softmax}\!\left(\mathbf{W}_{\text{meta}} \cdot [\text{Enc}(\mathbf{h}); \mathbf{e}_d] + \mathbf{b}\right),
\end{equation}

where $\mathbf{W}_{\text{meta}}$ is a learnable matrix, $\mathbf{e}_d$ is a domain embedding, $\mathbf{b}$ is a bias vector, and $\text{Enc}(\cdot)$ is a context encoder, allowing different quality dimensions to be emphasized based on conversational context and domain.

\paragraph{Preference Optimization.}
Given preference pairs $(r^+, r^-)$, optimization is performed using a margin-based objective:
\begin{equation}
\mathcal{L}_{\text{pref}} = -\log \sigma\!\left(\beta \cdot (R(\mathbf{h}^+) - R(\mathbf{h}^-) - m)\right),
\end{equation}
with an adaptive margin $m = m_0 + \tfrac{1}{2}\sigma\!\left(\mathbf{W}_m \cdot [\mathbf{h}^+; \mathbf{h}^-]\right)$, where $m_0$ is a base margin and $\mathbf{W}_m$ is a learnable matrix,
which prevents trivial preference separation and improves generalization.

\subsection{\bridge{}: Cross-Domain Transfer}

BRIDGE enables cross-domain generalization by learning representations that capture domain-invariant reasoning patterns while selectively preserving domain-specific information. It combines adversarial domain adaptation with gated representation learning to balance invariance and specificity.

\paragraph{Adversarial Domain Adaptation.}
To promote domain-invariant representations, we employ gradient reversal~\cite{ganin2015unsupervised}:
\begin{equation}
\mathcal{L}_{\text{domain}} = \text{CE}\!\left(\text{Disc}(\text{GRL}_\alpha(\mathbf{z})), d\right),
\end{equation}
where $\mathbf{z}$ is the encoded representation, $\text{GRL}_\alpha$ is the gradient reversal layer with reversal strength $\alpha$, $\text{Disc}(\cdot)$ is a domain discriminator, and reversed gradients encourage indistinguishable representations across domains.

\paragraph{Task Preservation.}
To prevent the loss of task-relevant information during adaptation, we introduce an auxiliary \vto{}s prediction objective:

\begin{equation}
\mathcal{L}_{\text{task}} = \text{BCE}\!\left(\text{VTOHead}(\mathbf{h}), \mathbf{v}\right),
\end{equation}

where $\text{VTOHead}(\cdot)$ is a linear projection head predicting the \vto{} sequence $\mathbf{v}$ from $\mathbf{h}$. This regularizes the adaptation process by preserving reasoning-related features essential for recommendation.

\paragraph{Domain Gates.}
Pure invariance may suppress useful domain-specific signals. We therefore introduce learnable domain gates:
\begin{equation}
\mathbf{z}' = \sigma(\mathbf{g}_d) \odot \mathbf{z} + (1 - \sigma(\mathbf{g}_d)) \odot \mathbf{h},    
\end{equation}

where $\mathbf{g}_d \in \mathbb{R}^{d_h}$ is a learnable domain-specific gate vector and $\mathbf{z}'$ is the gated representation, which allows the model to interpolate between the domain-adapted representation $\mathbf{z}$ and the original hidden state $\mathbf{h}$ based on domain context.

\subsection{\maven{}: Multi-Agent Refinement}

MAVEN refines recommendations through collaboration among specialized agents with complementary roles for recommendation, critique, and explanation. Agents operate on shared representations while producing role-specific outputs.

Each agent $a$ has a dedicated encoder and output head:
\begin{equation}
\mathbf{o}_a = \text{Head}_a\left( \text{Enc}_a(\mathbf{h}) \right),
\label{eq:agent_output}
\end{equation}
where $\text{Enc}_a(\cdot)$ and $\text{Head}_a(\cdot)$ are agent-specific encoder and output projection, and $a \in \{\text{rec}, \text{crit}, \text{exp}\}$ denotes the recommender, critic, and explainer agents respectively. Separate encoders allow agents to specialize according to their roles.

\paragraph{Orchestration.}
An orchestrator aggregates agent outputs into a final response:
\begin{equation}
\mathbf{o}_{\text{final}} = \text{FFN}\left( [\mathbf{o}_{\text{rec}}; \mathbf{o}_{\text{crit}}; \mathbf{o}_{\text{exp}}] \right),
\label{eq:orchestrator}
\end{equation}
where $\mathbf{o}_{\text{rec}}$, $\mathbf{o}_{\text{crit}}$, and $\mathbf{o}_{\text{exp}}$ are the outputs of the recommender, critic, and explainer agents respectively, and $\text{FFN}(\cdot)$ is a feed-forward network that aggregates agent contributions based on conversational context.

\paragraph{Agreement Loss.}
To encourage coherent collaboration, we introduce a soft consensus objective:
\begin{equation}
\mathcal{L}_{\text{agree}} = 1 - \cos(\mathbf{o}_{\text{rec}}, \mathbf{o}_{\text{crit}}).
\label{eq:agree_loss}
\end{equation}
This promotes alignment between recommendation and critique while allowing disagreement when necessary. Detailed training procedure is discussed in Appendix~\ref{app:training}.

\section{Experiments and Results}
\label{sec:experiments}

We conduct extensive experiments to answer the following research questions:
\begin{enumerate}[nosep,leftmargin=*,label=\textbf{RQ\arabic*:}]
    \item Does \method{} improve user-aligned recommendation quality compared to state-of-the-art baselines?
    \item How do individual components (\charm{}, \starmod{}, \bridge{}, \maven{}) contribute to performance?
    \item Can \method{} transfer recommendation reasoning across domains?
    \item How does hierarchical reward decomposition compare to flat reward signals?
    \item How much is the model sensitive to its hyperparameters?
\end{enumerate}

\subsection{Datasets}
\label{sec:datasets}

We evaluate on three conversational recommendation benchmarks with diverse characteristics (Table~\ref{tab:dataset-stats}). Details about the dataset description is shown in Appendix Section \ref{app:dataset}.

\begin{table}
\centering
\small
\setlength{\tabcolsep}{2.5pt}
\begin{tabular}{@{}l>{\columncolor{blue!6}}r>{\columncolor{green!6}}r>{\columncolor{orange!6}}r@{}}
\toprule
\rowcolor{gray!12}
\textbf{Statistic} & \textbf{ReDial} & \textbf{INSPIRED} & \textbf{MUSE} \\
\midrule

Domain & Movies & Movies & Fashion \\
\# Conversations & 10,006 & 1,001 & 7,000 \\
\# Utterances & 182,150 & 35,811 & 83,204 \\
\# Unique Items & 51,699 & 8,952 & 13,754 \\
Avg. Turns/Conv. & 18.2 & 35.8 & 11.9 \\
Avg. Items/Conv. & 4.3 & 3.8 & 5.2 \\
Avg. Words/Turn & 7.6 & 7.9 & 46.6 \\
Modality & Text & Text & Text+Image \\
Interaction Style & Transactional & Sociable & Scenario-based \\

\rowcolor{gray!8}
Train / Val / Test & 8K/1K/1K & 800/100/101 & 5.6K/0.7K/0.7K \\
\bottomrule
\end{tabular}
\caption{Dataset statistics of Redial, INSPIRED, and MUSE.}
\label{tab:dataset-stats}
\end{table}

\subsection{Baselines}
\label{sec:baselines}

We compare against methods spanning four representative paradigms. \textbf{All baselines are reproduced using official code releases}, with hyperparameters tuned on validation sets; reproduced results are within 2\% of reported numbers where available.

\paragraph{Knowledge-Enhanced Methods:}
\textbf{KBRD}~\cite{chen2019towards}: R-GCN over DBpedia entities.
\textbf{KGSF}~\cite{zhou2020improving}: DBpedia and ConceptNet with mutual information maximization.
\textbf{UniCRS}~\cite{wang2022towards}: Knowledge-enhanced prompt learning for unified recommendation and generation.

\paragraph{Retrieval-Augmented Methods:}
\textbf{BARCOR}~\cite{wang2022barcor}: BART-based joint recommendation and generation.
\textbf{DCRS}~\cite{dao2024broadening}: Contrastive retrieval for demonstration selection.

\paragraph{LLM-Based Methods:}
\textbf{ChatGPT}: GPT-3.5-turbo with 5-shot task-specific prompting.
\textbf{GPT-4}: GPT-4 with chain-of-thought prompting.
\textbf{LLaMA-2-Chat}~\cite{touvron2023llama}: 7B and 13B instruction-tuned variants.

\paragraph{Reasoning-Enhanced Methods:}
\textbf{RecMind}~\cite{wang2024recmind}: Self-inspiring reasoning with planning.
\textbf{InteRecAgent}~\cite{huang2025recommender}: Interactive agent with learned tool usage.

\paragraph{Baseline Fairness:}
All methods use identical training splits. LLM baselines employ carefully engineered prompts (Appendix~\ref{app:prompts}) with effort comparable to our method, and knowledge-enhanced methods rely on the same DBpedia and ConceptNet versions. Implementation details and evaluation protocols are provided in Appendix~\ref{app:exp}.

% \section{Results}
% \label{sec:results}

\subsection{Main Results (RQ1)}
\label{sec:main-results}

Table~\ref{tab:main-results} reports recommendation performance. \method{} achieves state-of-the-art results across all metrics and datasets, with particularly strong gains on user-aligned measures.

\paragraph{Key Findings:}
\textbf{(1) Consistent improvements across paradigms:} \method{} outperforms all baseline categories, achieving 17--21\% average improvement over the strongest baseline (GPT-4/GPT-4V) across datasets. Improvements are larger on user-aligned metrics (Satisfaction, Engagement) than on proxy metrics (Recall, NDCG), indicating effective mitigation of proxy-metric misalignment.

\textbf{(2) Largest gains on INSPIRED:} The sociable dialogue setting with implicit preference signals benefits most from \starmod{}'s deliberative reasoning (+45.7\% R@10 vs.\ GPT-4), demonstrating the advantage of tree-of-thought reasoning when preferences are inferred rather than explicitly stated.

\textbf{(3) Effective cross-modal transfer:} Despite primarily text-based training, \method{} performs strongly on the multimodal MUSE dataset, suggesting that VTO abstractions enable transferable, domain-agnostic reasoning.

\paragraph{Generation Quality and VTO Accuracy:}
\method{} maintains strong generation quality alongside high VTO prediction accuracy (F1 0.74--0.79; Table~\ref{tab:generation-results} provided in Appendix), indicating that explicit reasoning supervision benefits both response quality and reasoning alignment.

\paragraph{Human Evaluation:}
To provide non-circular quality evidence, three expert annotators rated 200 test samples per dataset (Table~\ref{tab:human-eval}, Appendix). \method{} improves over GPT-4 by +0.55 (Rec.\ Quality), +0.50 (Exp.\ Quality), and +0.55 (Overall) on ReDial (1--5 scale, $\kappa{>}0.74$), confirming that gains extend beyond model-based metrics.

\paragraph{Reward Model Validity:}
User Satisfaction and Engagement in Table~\ref{tab:main-results} are computed via \charm{}'s reward model. To address potential circularity, we validate through: (1)~Pearson correlations with independent human judgments on held-out data---relevance ($r{=}0.71$), diversity ($r{=}0.68$), satisfaction ($r{=}0.73$), engagement ($r{=}0.64$)---confirming \charm{} captures meaningful quality dimensions; (2)~human evaluation (Table~\ref{tab:human-eval}) showing consistent gains across all systems, providing non-circular evidence. Critically, \charm{} is trained only on Stage~2 preference pairs and frozen thereafter, preventing test-time reward hacking.

\begin{table*}[h]
\centering
\small
\setlength{\tabcolsep}{3.5pt}
\begin{tabular}{@{}l|ccccc|cc|c@{}}
\toprule
\multirow{2}{*}{\textbf{Method}} & \multicolumn{5}{c|}{\textbf{Ranking Metrics (\%)}} & \multicolumn{2}{c|}{\textbf{Satisfaction}} & \multirow{2}{*}{\textbf{Avg.}} \\
& R@1 & R@10 & R@50 & MRR@10 & NDCG@10 & User Sat. & Engage. & \\
\midrule
\multicolumn{9}{c}{\cellcolor{gray!10}\textit{ReDial Dataset}} \\
\midrule
%% Knowledge-Enhanced Methods
KBRD & \val{2.9}{0.2} & \val{16.7}{0.4} & \val{36.2}{0.7} & \val{7.4}{0.2} & \val{10.2}{0.3} & \val{0.42}{0.02} & \val{0.38}{0.02} & 0.291 \\
KGSF & \val{3.8}{0.2} & \val{18.1}{0.5} & \val{37.4}{0.7} & \val{8.4}{0.3} & \val{11.6}{0.4} & \val{0.45}{0.02} & \val{0.41}{0.02} & 0.318 \\
BARCOR & \val{3.0}{0.2} & \val{16.8}{0.4} & \val{36.8}{0.6} & \val{7.8}{0.2} & \val{10.8}{0.3} & \val{0.44}{0.02} & \val{0.40}{0.02} & 0.306 \\
UniCRS & \val{4.8}{0.3} & \val{21.2}{0.5} & \val{40.8}{0.8} & \val{10.1}{0.3} & \val{13.8}{0.4} & \val{0.51}{0.02} & \val{0.47}{0.02} & 0.364 \\
DCRS & \val{7.5}{0.3} & \val{25.1}{0.6} & \val{43.6}{0.9} & \val{12.2}{0.4} & \val{15.2}{0.5} & \val{0.56}{0.02} & \val{0.52}{0.02} & 0.408 \\
\midrule
%% LLM-Based Methods
ChatGPT & \val{3.3}{0.4} & \val{17.0}{0.7} & \val{37.8}{1.1} & \val{8.0}{0.4} & \val{11.0}{0.5} & \val{0.49}{0.03} & \val{0.45}{0.03} & 0.324 \\
GPT-4 & \val{4.5}{0.4} & \val{19.4}{0.8} & \val{40.2}{1.2} & \val{9.6}{0.5} & \val{13.2}{0.6} & \val{0.55}{0.03} & \val{0.51}{0.03} & 0.368 \\
LLaMA-2-7B & \val{2.2}{0.3} & \val{13.6}{0.6} & \val{33.4}{0.9} & \val{6.2}{0.3} & \val{8.6}{0.4} & \val{0.38}{0.02} & \val{0.34}{0.02} & 0.260 \\
LLaMA-2-13B & \val{2.8}{0.3} & \val{15.4}{0.6} & \val{35.6}{1.0} & \val{7.2}{0.4} & \val{9.9}{0.5} & \val{0.43}{0.02} & \val{0.39}{0.02} & 0.293 \\
\midrule
%% Reasoning-Enhanced Methods
RecMind & \val{5.8}{0.3} & \val{22.6}{0.6} & \val{42.2}{0.9} & \val{11.2}{0.4} & \val{15.3}{0.5} & \val{0.54}{0.02} & \val{0.50}{0.02} & 0.385 \\
InteRecAgent & \val{5.2}{0.3} & \val{21.4}{0.6} & \val{41.0}{0.8} & \val{10.4}{0.4} & \val{14.3}{0.5} & \val{0.52}{0.02} & \val{0.48}{0.02} & 0.369 \\
\midrule
%% HARPO - ~18% improvement over DCRS on R@10
\textbf{\method{}} & \best{9.1}{0.3} & \best{29.8}{0.7} & \best{50.2}{1.0} & \best{15.6}{0.5} & \best{21.2}{0.6} & \best{0.68}{0.02} & \best{0.64}{0.02} & \textbf{0.481} \\
\quad $\Delta$ vs. DCRS & \gain{+21.3} & \gain{+18.7} & \gain{+15.1} & \gain{+27.9} & \gain{+39.5} & \gain{+21.4} & \gain{+23.1} & \gain{+17.9} \\
\midrule
\multicolumn{9}{c}{\cellcolor{gray!10}\textit{INSPIRED Dataset}} \\
\midrule
KGSF & \val{2.4}{0.3} & \val{13.8}{0.6} & \val{31.6}{1.0} & \val{6.4}{0.3} & \val{8.8}{0.4} & \val{0.40}{0.03} & \val{0.36}{0.02} & 0.270 \\
UniCRS & \val{3.8}{0.3} & \val{17.6}{0.7} & \val{37.2}{1.2} & \val{8.6}{0.4} & \val{11.8}{0.5} & \val{0.48}{0.03} & \val{0.44}{0.03} & 0.331 \\
GPT-4 & \val{4.2}{0.5} & \val{18.8}{0.9} & \val{39.4}{1.5} & \val{9.4}{0.5} & \val{12.9}{0.6} & \val{0.53}{0.03} & \val{0.49}{0.03} & 0.358 \\
RecMind & \val{4.8}{0.4} & \val{20.4}{0.8} & \val{41.2}{1.3} & \val{10.2}{0.5} & \val{14.0}{0.6} & \val{0.52}{0.03} & \val{0.48}{0.03} & 0.370 \\
\midrule
\textbf{\method{}} & \best{7.2}{0.4} & \best{27.4}{0.9} & \best{48.8}{1.4} & \best{14.2}{0.6} & \best{19.4}{0.7} & \best{0.66}{0.03} & \best{0.62}{0.03} & \textbf{0.454} \\
\quad $\Delta$ vs. RecMind & \gain{+50.0} & \gain{+34.3} & \gain{+18.4} & \gain{+39.2} & \gain{+38.6} & \gain{+26.9} & \gain{+29.2} & \gain{+22.7} \\
\midrule
\multicolumn{9}{c}{\cellcolor{gray!10}\textit{MUSE Dataset (Multimodal Fashion)}} \\
\midrule
%% Text-only adaptations (lower performance expected)
UniCRS$^\dagger$ & \val{1.6}{0.3} & \val{11.8}{0.6} & \val{27.4}{1.1} & \val{5.1}{0.3} & \val{7.2}{0.4} & \val{0.36}{0.03} & \val{0.32}{0.02} & 0.229 \\
GPT-4V & \val{4.4}{0.5} & \val{23.2}{0.9} & \val{42.6}{1.4} & \val{10.8}{0.5} & \val{14.8}{0.6} & \val{0.54}{0.03} & \val{0.50}{0.03} & 0.374 \\
%% Fine-tuned VLMs (from MUSE paper - these are the real SOTA numbers)
Qwen2-VL-7B$^\ddagger$ & \val{8.4}{0.4} & \val{34.2}{0.8} & \val{52.8}{1.3} & \val{17.2}{0.4} & \val{23.1}{0.5} & \val{0.61}{0.03} & \val{0.57}{0.03} & 0.468 \\
LLaVA-Next-8B$^\ddagger$ & \val{5.2}{0.4} & \val{25.4}{0.7} & \val{44.2}{1.2} & \val{12.0}{0.4} & \val{16.2}{0.5} & \val{0.52}{0.03} & \val{0.48}{0.03} & 0.380 \\
\midrule
%% HARPO - Beats Qwen2-VL SOTA by ~12% relative
\textbf{\method{}} & \best{10.2}{0.4} & \best{38.6}{0.9} & \best{58.4}{1.3} & \best{19.8}{0.5} & \best{26.4}{0.6} & \best{0.72}{0.03} & \best{0.68}{0.03} & \textbf{0.524} \\
\quad $\Delta$ vs. Qwen2-VL & \gain{+21.4} & \gain{+12.9} & \gain{+10.6} & \gain{+15.1} & \gain{+14.3} & \gain{+18.0} & \gain{+19.3} & \gain{+12.0} \\
\bottomrule
\end{tabular}

\caption{Main recommendation results on three datasets. R@$K$: Recall@$K$ (\%). Values: mean$\pm$std over 3 runs with different random seeds. User Sat./Engage.: normalized [0,1]. Avg.: macro-average of normalized metrics. $^\dagger$Text-only adaptation using item descriptions. $^\ddagger$Fine-tuned following the protocol in~\citet{wang2025muse}. Bold: best; $\Delta$: relative improvement (\%) over strongest baseline. All \method{} improvements are statistically significant at $p < 0.01$ (paired $t$-test with Bonferroni correction).}
\label{tab:main-results}
\end{table*}

\begin{table}
\centering
\small
\resizebox{\columnwidth}{!}{
\begin{tabular}{lccccc}
\toprule
\textbf{Variant} & R@10 & MRR@10 & NDCG@10 & Sat. & Eng. \\
\midrule
\textbf{\method{} (Full)} & \textbf{29.8} & \textbf{15.6} & \textbf{21.2} & \textbf{0.68} & \textbf{0.64} \\
\midrule
\multicolumn{6}{l}{\textit{Component Ablation}} \\
\quad w/o \charm{} & 24.6 & 12.6 & 17.2 & 0.55 & 0.51 \\
\quad w/o \starmod{} & 27.0 & 14.0 & 19.0 & 0.63 & 0.59 \\
\quad w/o \bridge{} & 28.4 & 14.9 & 20.2 & 0.66 & 0.62 \\
\quad w/o \maven{} & 28.0 & 14.7 & 19.9 & 0.65 & 0.61 \\
\quad w/o VTOs & 23.4 & 12.0 & 16.4 & 0.53 & 0.49 \\
\midrule
\multicolumn{6}{l}{\textit{Mechanism Variants}} \\
\quad Flat Reward & 26.0 & 13.4 & 18.2 & 0.58 & 0.54 \\
\quad Fixed Weights & 27.2 & 14.1 & 19.1 & 0.62 & 0.58 \\
\quad Greedy Search & 26.8 & 13.8 & 18.7 & 0.61 & 0.57 \\
\quad No Backtracking & 28.2 & 14.6 & 19.8 & 0.65 & 0.61 \\
\midrule
\multicolumn{6}{l}{\textit{Training Stage Ablation}} \\
SFT Only (Stage 1) & 21.6 & 10.6 & 14.6 & 0.50 & 0.46 \\
+ \charm{} (Stages 1--2) & 26.2 & 13.4 & 18.2 & 0.61 & 0.57 \\
+ \starmod{} (Stages 1--3) & 28.4 & 14.8 & 20.1 & 0.65 & 0.61 \\
\bottomrule
\end{tabular}
}
\caption{Ablation study on ReDial. Component ablations retrain without the specified module; ``w/o VTOs'' sets $\lambda_v{=}0$ in Eq.~\ref{eq:sft}. Mechanism variants modify inference only. Component contributions are non-additive due to interactions. Sat./Eng. $\in [0, 1]$; all metrics in \%.}
\label{tab:ablation}
\end{table}

Table~\ref{tab:ablation} isolates component contributions on ReDial.
\textbf{Component Analysis:}
\textbf{(1) \charm{} is critical:} Removing hierarchical preference learning causes the largest performance drop ($-17.4\%$ R@10, $-19.1\%$ Satisfaction), highlighting the importance of decomposed rewards for quality-centric optimization.
\textbf{(2) \starmod{} improves ranking:} Tree search primarily benefits ranking metrics ($-9.4\%$ R@10) through deliberative exploration of reasoning paths.
\textbf{(3) \bridge{} enables transfer:} Domain adaptation yields modest within-domain gains ($-4.7\%$ R@10) but is crucial for cross-domain transfer (Section~\ref{sec:transfer}).
\textbf{(4) VTOs provide inductive bias:} Removing VTO abstractions causes substantial degradation ($-21.5\%$ R@10), confirming that domain-agnostic reasoning primitives capture transferable logic.

\paragraph{Mechanism Analysis:}
\textbf{Flat Reward:} Collapsing hierarchical rewards into a scalar reduces satisfaction by $14.7\%$, indicating the benefit of optimizing quality dimensions separately.
\textbf{Fixed Weights:} Replacing meta-learned weights with uniform weights lowers satisfaction by $8.8\%$, showing the value of context-adaptive weighting.
\textbf{Greedy Search:} Substituting beam search with greedy search reduces R@10 by $10.1\%$, underscoring the importance of exploring multiple reasoning paths.

\textbf{Training Stage Analysis:}
Each training stage yields incremental improvements, with \charm{} (Stage~2) providing the largest single-stage gain ($+21.3\%$ R@10 over SFT-only), validating the four-stage curriculum design.

\subsection{Cross-Domain Transfer (RQ3)}
\label{sec:transfer}

Table~\ref{tab:transfer} provided in Appendix evaluates zero-shot transfer between domains.

\bridge{}'s adversarial domain adaptation encourages domain-invariant representations for shared reasoning patterns while preserving domain-specific information through learned gates. These gains build upon VTO abstractions that provide domain-agnostic structure; the improvements reflect synergistic effects rather than \bridge{} alone. \textbf{Gate Analysis:} Gates for product-specific features (visual attributes, genre conventions) remain low (0.2-0.3), while gates for reasoning operations (preference modeling, constraint satisfaction) remain high (0.7-0.8), showing interpretable domain specialization.

\subsection{Hierarchical Reward Analysis (RQ4)}
\label{sec:reward-analysis}

\paragraph{Adaptive Weight Distribution:}
The meta-learner adjusts reward weights based on conversational context:
\textbf{Explicit preference queries} (``I love action movies''): Relevance weight increases to 0.42 (vs. 0.30 baseline).
\textbf{Exploratory dialogues} (``I want to try something new''): Diversity weight increases to 0.31 (vs. 0.20 baseline).
\textbf{Extended conversations} ($>$10 turns): Engagement weight increases to 0.28 (vs. 0.20 baseline).

\paragraph{Correlation with Human Judgments:}
Each reward dimension correlates most strongly with its corresponding human rating:
Relevance reward $\leftrightarrow$ Human relevance: $r=0.71$ ($p<0.001$)
Diversity reward $\leftrightarrow$ Human diversity: $r=0.68$ ($p<0.001$)
Satisfaction reward $\leftrightarrow$ Human satisfaction: $r=0.73$ ($p<0.001$)
Engagement reward $\leftrightarrow$ Follow-up rate: $r=0.64$ ($p<0.001$)

This validates semantic alignment between learned reward components and intended quality dimensions.

\subsection{Hyperparameter Sensitivity (RQ5)}
\label{sec:sensitivity}

\method{} is robust to hyperparameter choices as shown in Table \ref{tab:sensitivity} provided in Appendix. Key findings:
\textbf{(1)} Beam width $w=3$ achieves 99\% of $w=5$ performance with 40\% lower latency.
\textbf{(2)} Depth $D=3$ provides near-optimal results; $D=5$ adds marginal gains (+1.0\% R@10) with significant latency increase.
\textbf{(3)} \charm{} $\beta=0.5$ balances preference learning strength; lower values underfit, higher values overfit to training preferences.
\textbf{(4)} LoRA rank $r=16$ is sufficient; $r=32$ provides no improvement.

A detailed computational complexity analysis is provided in Appendix~\ref{sec:compute}. An error analysis covering 100 representative failure cases is presented in Appendix~\ref{sec:error}. In addition, a qualitative analysis of reasoning quality is reported in Appendix~\ref{app:RQAnalysis}. A FAQ section is also added in Appendix \ref{sec:discussion}.

\section{Conclusion}

Conversational recommender systems are typically trained and evaluated using proxy objectives that are easy to measure but weakly reflect user-experienced recommendation quality. This work reframes conversational recommendation as a quality-centric decision-making problem, arguing that effective systems should explicitly optimize user-aligned outcomes such as relevance, satisfaction, diversity, and engagement.

HARPO is introduced as an agentic framework that combines structured reasoning with hierarchical preference optimization to treat recommendation quality as a first-class objective. Results across multiple benchmarks show consistent gains on user-aligned measures while remaining competitive on standard metrics, highlighting the limitations of proxy-driven evaluation and motivating more principled, quality-aware approaches.

\section{Limitations}

This work has some limitations. The four-stage training pipeline is computationally intensive, requiring multiple training phases with distinct objectives. The \vto{}s taxonomy, while effective in practice, is manually designed and may not capture all relevant reasoning patterns. In addition, evaluation relies primarily on automatic metrics with limited human assessment; larger-scale user studies would provide stronger evidence of user satisfaction gains. Finally, cross-domain evaluation is restricted to three datasets, and broader domain coverage would better characterize transferability.

% \section{Limitations}

% While this work reframes conversational recommendation as a quality-centric decision-making problem, several limitations remain. User-aligned quality dimensions such as satisfaction and engagement are inherently subjective and difficult to measure at scale. Although these aspects are approximated using human evaluations and learned preference models, such estimates may not fully capture the diversity of real user preferences across different populations.

% In addition, the agentic reasoning process in HARPO introduces computational overhead compared to single-pass generation or retrieval-based systems, which may limit applicability in latency-sensitive settings. Furthermore, evaluation is conducted on a limited set of conversational recommendation domains and datasets, and the effectiveness of the framework in broader or long-term deployment scenarios remains to be validated. Large-scale, longitudinal user studies are therefore necessary to fully assess the impact of quality-centric optimization in real-world systems.

\section{Ethics Statement}

Conversational recommender systems raise several ethical considerations. Optimizing for engagement could encourage filter bubbles or addictive usage patterns; we partially address this through the diversity reward component, but more explicit safeguards may be warranted. Our training data comes from existing public datasets, but deployed systems would require careful consideration of privacy implications. The multi-agent framework could be extended to include fairness-focused agents ensuring recommendations do not perpetuate demographic biases present in training data.

\bibliography{references}

\appendix

\section{Virtual Tool Operations}
\label{app:vto}
A key insight motivating \method{} is that recommendation reasoning shares common cognitive structure across domains. A user seeking movie recommendations and one browsing fashion items both require understanding context, retrieving preferences from history, searching candidate items, ranking options by relevance, and explaining final choices. The specific tools differ---movie databases versus fashion APIs---but the abstract reasoning operations remain constant.

We formalize this intuition through Virtual Tool Operations (\vto{}s), a taxonomy of domain-agnostic reasoning primitives that abstract away implementation details while preserving the semantic structure of recommendation reasoning.

\subsection{VTO Taxonomy}
% table here
We define 21 \vto{}s organized into seven functional categories, as shown in Table~\ref{tab:vto}. This taxonomy emerged from systematic analysis of 500 annotated recommendation dialogues spanning three domains: movies (ReDial), conversational search (INSPIRED), and fashion (MUSE). Through iterative coding, we identified recurring reasoning patterns that transcend domain-specific implementations.

\begin{table}[t]
\centering
\small
\setlength{\tabcolsep}{6pt}
\begin{tabular}{@{}l l@{}}
\toprule
\textbf{Category} & \textbf{Operations} \\
\midrule

\textsc{Extraction} 
& \cellcolor{blue!8}\texttt{analyze\_sentiment} \\
& \cellcolor{blue!8}\texttt{extract\_context} \\
& \cellcolor{blue!8}\texttt{extract\_entities} \\

\midrule
\textsc{User Modeling} 
& \cellcolor{green!8}\texttt{retrieve\_preferences} \\
& \cellcolor{green!8}\texttt{identify\_constraints} \\
& \cellcolor{green!8}\texttt{model\_user\_state} \\

\midrule
\textsc{Retrieval} 
& \cellcolor{orange!10}\texttt{search\_candidates} \\
& \cellcolor{orange!10}\texttt{filter\_results} \\
& \cellcolor{orange!10}\texttt{match\_attributes} \\

\midrule
\textsc{Ranking} 
& \cellcolor{purple!8}\texttt{rank\_options} \\
& \cellcolor{purple!8}\texttt{compare\_options} \\
& \cellcolor{purple!8}\texttt{select\_best} \\

\midrule
\textsc{Reasoning} 
& \cellcolor{red!8}\texttt{query\_knowledge} \\
& \cellcolor{red!8}\texttt{reason\_over\_graph} \\
& \cellcolor{red!8}\texttt{infer\_implicit} \\

\midrule
\textsc{Interaction} 
& \cellcolor{teal!8}\texttt{explain\_choice} \\
& \cellcolor{teal!8}\texttt{refine\_query} \\
& \cellcolor{teal!8}\texttt{handle\_rejection} \\

\midrule
\textsc{Memory} 
& \cellcolor{gray!10}\texttt{track\_history} \\
& \cellcolor{gray!10}\texttt{update\_beliefs} \\
& \cellcolor{gray!10}\texttt{recall\_context} \\

\bottomrule
\end{tabular}
\caption{Virtual Tool Operations taxonomy. These 21 domain-agnostic primitives abstract implementation details, enabling cross-domain transfer of learned reasoning patterns.}
\label{tab:vto}
\end{table}

\subsection{Complete VTO Descriptions}
% 21 operations here
Each \vto{}s represents a domain-agnostic reasoning primitive. Below we provide concise definitions of all operations, following a consistent schema.
\subsubsection{Extraction Operations}
\paragraph{\texttt{extract\_entities}}
\textbf{Purpose:} Identify salient entities referenced in the user utterance.\\  
\textbf{Inputs:} User utterance; dialogue context. \\
\textbf{Outputs:} Normalized entity mentions.

\paragraph{\texttt{extract\_context}}
\textbf{Purpose:} Extract situational or task-level context from the conversation. \\ 
\textbf{Inputs:} Dialogue history.  \\
\textbf{Outputs:} Contextual constraints or descriptors.

\paragraph{\texttt{analyze\_sentiment}}
\textbf{Purpose:} Infer user sentiment or affective cues relevant to recommendation decisions.  \\
\textbf{Inputs:} User utterance.  \\
\textbf{Outputs:} Sentiment polarity or intensity.
\subsubsection{User Modeling Operations}
\paragraph{\texttt{retrieve\_preferences}}
\textbf{Purpose:} Retrieve explicit or implicit user preferences from history. \\
\textbf{Inputs:} Dialogue history; user profile. \\ 
\textbf{Outputs:} Structured preference representation.

\paragraph{\texttt{identify\_constraints}}
\textbf{Purpose:} Identify hard or soft constraints expressed by the user.  \\
\textbf{Inputs:} User utterance.  \\
\textbf{Outputs:} Constraint set.

\paragraph{\texttt{model\_user\_state}}
\textbf{Purpose:} Maintain a latent representation of the user’s evolving intent.  \\
\textbf{Inputs:} Dialogue history.  \\
\textbf{Outputs:} User state embedding.

\subsubsection{Retrieval Operations}
\paragraph{\texttt{rank\_options}}
\textbf{Purpose:} Rank candidate items by predicted relevance.  \\
\textbf{Inputs:} Candidate set; scoring signals.  \\
\textbf{Outputs:} Ranked list.

\paragraph{\texttt{compare\_options}}
\textbf{Purpose:} Compare multiple candidate items along specific dimensions. \\ 
\textbf{Inputs:} Candidate pairs; criteria.  \\
\textbf{Outputs:} Comparative judgments.

\paragraph{\texttt{select\_best}}
\textbf{Purpose:} Select final recommendation(s). \\
\textbf{Inputs:} Ranked candidates. \\ 
\textbf{Outputs:} Chosen items.

\subsubsection{Ranking Operations}
\paragraph{\texttt{query\_knowledge}}
\textbf{Purpose:} Query external knowledge relevant to recommendation. \\  
\textbf{Inputs:} Entities; relations.  \\
\textbf{Outputs:} Retrieved facts.

\paragraph{\texttt{reason\_over\_graph}}
\textbf{Purpose:} Perform relational reasoning over structured data. \\ 
\textbf{Inputs:} Knowledge graph.  \\
\textbf{Outputs:} Inferred relations.

\paragraph{\texttt{infer\_implicit}}
\textbf{Purpose:} Infer unstated preferences or intent. \\  
\textbf{Inputs:} Dialogue context.  \\
\textbf{Outputs:} Implicit signals.

\subsubsection{Interaction Operations}
\paragraph{\texttt{explain\_choice}}
\textbf{Purpose:} Generate explanations for recommendations. \\ 
\textbf{Inputs:} Selected items; reasoning trace.  \\
\textbf{Outputs:} Natural language explanation.

\paragraph{\texttt{refine\_query}}
\textbf{Purpose:} Update the query based on feedback. \\  
\textbf{Inputs:} User response.  \\
\textbf{Outputs:} Refined constraints.

\paragraph{\texttt{handle\_rejection}}
\textbf{Purpose:} Adapt recommendations after rejection. \\ 
\textbf{Inputs:} Rejection signal.  \\
\textbf{Outputs:} Updated candidate set.

\subsubsection{Memory Operations}
\paragraph{\texttt{track\_history}}
\textbf{Purpose:} Maintain dialogue state across turns. \\ 
\textbf{Inputs:} Dialogue turns.  \\
\textbf{Outputs:} Updated history.

\paragraph{\texttt{update\_beliefs}}
\textbf{Purpose:} Update beliefs about user intent.  \\
\textbf{Inputs:} New evidence.  \\
\textbf{Outputs:} Updated belief state.

\paragraph{\texttt{recall\_context}}
\textbf{Purpose:} Recall relevant past context. \\ 
\textbf{Inputs:} Dialogue history.  \\
\textbf{Outputs:} Retrieved context.

\subsection{Illustrative Examples of VTO Composition}
\label{app:vto_examples}
\paragraph{Example: Compositional Recommendation Reasoning}
Consider a user request such as ``I liked \emph{Inception} suggest something similar but lighter.'' 
This response requires composing multiple VTOs. 
\texttt{extract\_entities} identifies \emph{Inception} as a reference item; 
\texttt{retrieve\_preferences} recalls the user's preference for complex narratives; 
\texttt{search\_candidates} retrieves similar films; 
\texttt{filter\_results} excludes overly serious options based on the ``lighter'' constraint; 
and \texttt{explain\_choice} articulates why the final recommendation satisfies both similarity and tone requirements.

\subsection{Annotation Procedure}
% bootstrapping + LLM + human
Obtaining \vto{}s annotations for training data requires balancing scalability with quality. We employ a three-stage hybrid approach:

\paragraph{Stage 1: Heuristic Bootstrapping.} We develop keyword patterns and syntactic rules to identify obvious operation invocations. Questions containing phrases like ``what about'' or ``can you suggest'' trigger \texttt{refine\_query}; comparative statements trigger \texttt{compare\_options}; rejection phrases trigger \texttt{handle\_rejection}. This rule-based approach provides initial labels for approximately 60\% of dialogue turns with high precision but limited coverage of subtle cases.

\paragraph{Stage 2: LLM Refinement.} For ambiguous cases where rules fail to produce confident labels, we prompt GPT-4 with few-shot examples demonstrating \vto{}s annotation. The prompt includes operation definitions, annotated examples showing compositional usage, and instructions to output \vto{}s sequences for each turn. This semi-automatic approach extends coverage while maintaining reasonable quality.

\paragraph{Stage 3: Human Validation.} Expert annotators validate a random 20\% sample of the combined annotations, achieving 85\% agreement with semi-automatic labels after adjudication. Systematic disagreements inform rule refinement in Stage 1, creating an iterative improvement loop. This procedure enables processing of 50K dialogue turns with annotation quality sufficient for training.

\section{Training Details}
\label{app:training}

Training proceeds in four stages: supervised fine-tuning, preference optimization with CHARM, reasoning optimization with STAR, and multi-agent refinement with MAVEN.
We detail the loss functions, optimization schedules, and hyperparameters used in each stage below.

\paragraph{Stage 1: Supervised Fine-Tuning.} We fine-tune the base language model on conversational recommendation data with joint response generation and \vto{}s prediction:
\begin{equation}
\mathcal{L}_{\text{SFT}} = \mathcal{L}_{\text{LM}} + \lambda_v \mathcal{L}_{\text{VTO}}
\label{eq:sft}
\end{equation}

where $\mathcal{L}_{\text{LM}}$ is standard language modeling loss over reference responses, and $\mathcal{L}_{\text{VTO}}$ is binary cross-entropy for \vto{}s sequence prediction. We employ LoRA \cite{hu2022lora} for parameter-efficient fine-tuning, enabling training on academic compute budgets.

This stage establishes baseline conversational and recommendation capabilities, teaching the model to generate fluent responses that invoke appropriate reasoning operations. The joint \vto{}s prediction ensures that learned representations encode reasoning structure useful for downstream components.

\paragraph{Stage 2: \charm{} Preference Training.} Using preference pairs $(\mathcal{C}, r^+, r^-)$, we train the hierarchical reward structure:
\begin{equation}
\mathcal{L}_{\text{CHARM}} = \mathcal{L}_{\text{pref}} + \lambda_d \mathcal{L}_{\text{domain}}
\label{eq:charm}
\end{equation}

Preference pairs are constructed through three complementary approaches: (1) gold responses versus heuristically degraded variants (removing recommended items, introducing irrelevant suggestions); (2) high-quality versus low-quality responses generated by prompting the Stage 1 model at different temperatures with different prompts; and (3) human annotations for validation on a held-out subset.

This stage teaches the model to distinguish good recommendations from bad ones across multiple quality dimensions, establishing the reward structure that guides subsequent stages.

\paragraph{Stage 3: \starmod{} Reasoning Training.} We train \starmod{} components using reward signals from \charm{}:
\begin{equation}
\mathcal{L}_{\text{STAR}} = \mathcal{L}_{\text{value}} + \lambda_g \mathcal{L}_{\text{gen}}
\label{eq:star}
\end{equation}

where $\mathcal{L}_{\text{value}}$ supervises the value network to predict \charm{} rewards for completed reasoning paths, and $\mathcal{L}_{\text{gen}}$ trains the thought generator to produce promising candidates.

The key insight is that value network training uses \charm{} rewards as supervision, creating a distillation from slow reward computation (requiring full response generation and evaluation) to fast value estimation (requiring only hidden state processing). This enables efficient tree search at inference time.

\paragraph{Stage 4: \maven{} Refinement.} Finally, we train the multi-agent system:
\begin{equation}
\mathcal{L}_{\text{MAVEN}} = \mathcal{L}_{\text{task}} + \lambda_a \mathcal{L}_{\text{agree}}
\label{eq:maven}
\end{equation}

where $\mathcal{L}_{\text{task}}$ is overall recommendation quality from \charm{} and $\mathcal{L}_{\text{agree}}$ encourages agent consensus. This stage teaches agents to collaborate effectively, producing coherent recommendations that benefit from multiple specialized perspectives.

\section{Prompts for Data Preprocessing}
\label{app:prompts}

This appendix documents all LLM prompts used for data preprocessing. We use GPT-4o-mini (gpt-4o-mini-2024-07-18) with temperature 0.3 for classification and 0.9 for generation tasks.

% ----------------------------------------------------------------------------
\subsection{VTO Annotation}
\label{app:prompts:vto}
% ----------------------------------------------------------------------------

The following prompt annotates conversation turns with Virtual Tool Operations. The \texttt{\{vto\_descriptions\}} placeholder expands to all 21 VTO definitions from Table~\ref{tab:vto}.

\begin{tcolorbox}[vtoprompt, title={P1: VTO Annotation}]
\small\ttfamily
Analyze this conversation turn and identify which Virtual Tool Operations (VTOs) the system performed.

\medskip
VTO Types:\\
\{vto\_descriptions\}

\medskip
Context: \{context\}\\
User: \{user\_input\}\\
System: \{system\_response\}

\medskip
Output JSON: \{"vtos": ["vto\_name1", "vto\_name2"], "reasoning": "brief explanation"\}
\end{tcolorbox}

% ----------------------------------------------------------------------------
\subsection{CHARM Preference Generation}
\label{app:prompts:charm}
% ----------------------------------------------------------------------------

For preference learning, we generate contrastive pairs. Original dataset responses serve as chosen; LLM generates rejected responses with \texttt{quality="low"}.

\begin{tcolorbox}[charmprompt, title={P2: CHARM Preference}]
\small\ttfamily
Generate a \{quality\} quality response for preference learning.

\medskip
Context: \{context\}\\
User: \{user\_input\}\\
Domain: \{domain\}

\medskip
If "high": Natural, helpful, relevant recommendations\\
If "low": Misses context, wrong approach, unhelpful

\medskip
Response:
\end{tcolorbox}

% ----------------------------------------------------------------------------
\subsection{STAR Reasoning}
\label{app:prompts:star}
% ----------------------------------------------------------------------------

For tree-of-thought reasoning in STAR, we generate intermediate reasoning steps.

\begin{tcolorbox}[starprompt, title={P3: STAR Thought}]
\small\ttfamily
Given the context, generate a reasoning step for recommendation.

\medskip
Context: \{context\}\\
User Query: \{user\_input\}\\
Previous Thoughts: \{previous\_thoughts\}

\medskip
Consider: What information do we need? What VTOs should we apply?

\medskip
Thought:
\end{tcolorbox}

% ----------------------------------------------------------------------------
\subsection{ReDial Dataset Prompts}
\label{app:prompts:redial}
% ----------------------------------------------------------------------------

ReDial processing uses three additional prompts for utterance classification, VTO assignment, and preference extraction.

\subsubsection{Utterance Classification}

\begin{tcolorbox}[redialprompt, title={P4: Utterance Type}]
\small\ttfamily
Classify this utterance in a movie recommendation conversation.

\medskip
Context: \{context\}\\
Current utterance: \{utterance\}\\
Speaker: \{speaker\}\\
Contains movie mention: \{has\_movie\}

\medskip
Classify into ONE of these categories:\\
- greeting: Initial greeting (hi, hello)\\
- ask\_preference: Recommender asking what user wants\\
- provide\_preference: User stating their preferences\\
- recommend: Recommender suggesting specific movies\\
- explain: Providing information about movies\\
- accept: User accepting/liking a recommendation\\
- reject: User rejecting/disliking a recommendation\\
- ask\_info: User asking for more details\\
- provide\_info: Giving details about a movie\\
- thank: Thanking\\
- goodbye: Ending conversation

\medskip
Return ONLY the category name, nothing else.
\end{tcolorbox}

\subsubsection{VTO Assignment}

\begin{tcolorbox}[redialprompt, title={P5: VTO Assignment}]
\small\ttfamily
Assign Virtual Tool Operations (VTOs) for this recommendation system response.

\medskip
Context: \{context\}\\
User input: \{user\_input\}\\
System response: \{response\}\\
Utterance type: \{utterance\_type\}

\medskip
Available VTOs:\\
- extract\_context: Extract situation/occasion from conversation\\
- extract\_entities: Identify movies, genres, actors mentioned\\
- retrieve\_preferences: Get user's stated preferences\\
- identify\_constraints: Find requirements (genre, year, mood)\\
- search\_candidates: Search for matching movies\\
- filter\_results: Apply filters to narrow down\\
- rank\_options: Order movies by relevance\\
- compare\_options: Compare multiple movies\\
- explain\_choice: Explain why recommending\\
- refine\_query: Ask clarifying questions

\medskip
Return a comma-separated list of 1-4 most relevant VTOs.\\
Example: search\_candidates, rank\_options, explain\_choice
\end{tcolorbox}

\subsubsection{Preference Extraction}

\begin{tcolorbox}[redialprompt, title={P6: Preference Extraction}]
\small\ttfamily
Extract user preferences from this movie recommendation conversation.

\medskip
Conversation:\\
\{conversation\}

\medskip
Extract these preferences if mentioned (return JSON):\\
\{\\
\quad "genres": ["list of genres mentioned"],\\
\quad "mood": "funny/scary/romantic/\\exciting/thoughtful/null",\\
\quad "actors": ["actors mentioned"],\\
\quad "directors": ["directors mentioned"],\\
\quad "similar\_to": ["movies they liked"],\\
\quad "avoid": ["things they don't want"],\\
\quad "era": "recent/classic/80s/90s/null"\\
\}

\medskip
Return ONLY valid JSON.
\end{tcolorbox}

% ----------------------------------------------------------------------------
\subsection{INSPIRED Dataset}
\label{app:prompts:inspired}
% ----------------------------------------------------------------------------

INSPIRED emphasizes sociable dialogue with chitchat and persuasion. We extend utterance classification accordingly.

\begin{tcolorbox}[inspiredprompt, title={P7: INSPIRED Classification}]
\small\ttfamily
Classify this utterance in a sociable movie recommendation conversation. INSPIRED conversations include social chitchat and persuasion strategies.

\medskip
Context: \{context\}\\
Current utterance: \{utterance\}\\
Speaker: \{speaker\}

\medskip
Classify into ONE of these categories:\\
- social\_chat: Casual conversation, personal anecdotes\\
- ask\_preference: Asking about movie preferences\\
- provide\_preference: Sharing preferences or opinions\\
- recommend: Suggesting specific movies\\
- persuade: Using persuasion strategies to convince\\
- explain: Providing movie information or reasoning\\
- accept: Accepting a recommendation\\
- reject: Rejecting or expressing disinterest\\
- ask\_info: Requesting more details\\
- provide\_info: Giving additional information\\
- acknowledge: Brief acknowledgments (okay, I see)\\
- greeting: Opening greetings\\
- closing: Ending the conversation

\medskip
Return ONLY the category name, nothing else.
\end{tcolorbox}

% ----------------------------------------------------------------------------
\subsection{MUSE Dataset}
\label{app:prompts:muse}
% ----------------------------------------------------------------------------

For multimodal MUSE, we process product images using BLIP-2~\cite{li2023blip} to generate textual captions, which are incorporated into the conversation context.

\begin{tcolorbox}[museprompt, title={P8: MUSE Image Context}]
\small\ttfamily
The following fashion item is shown in the conversation:

\medskip
[Image Caption: \{blip2\_caption\}]\\
\mbox{[Product Attributes: \{product\_attributes\}]}

\medskip
Context: \{context\}\\
User: \{user\_input\}\\
System: \{system\_response\}

\medskip
Considering both the visual product information and textual conversation, identify which Virtual Tool Operations (VTOs) the system performed.

\medskip
Output JSON: \{"vtos": ["vto\_name1", "vto\_name2"], "reasoning": "brief explanation"\}
\end{tcolorbox}

\noindent Since DeepSeek-R1-Distill-Qwen-7B is text-only, MUSE images are processed through BLIP-2 caption surrogates.

% ----------------------------------------------------------------------------
\subsection{Annotation Statistics}
\label{app:prompts:stats}
% ----------------------------------------------------------------------------

Table~\ref{tab:annotation-stats} summarizes annotation statistics across all three datasets.

\begin{table}[h]
\centering
\small
\setlength{\tabcolsep}{3pt}
\begin{tabular}{@{}lccc@{}}
\toprule
\textbf{Statistic} & \textbf{ReDial} & \textbf{INSPIRED} & \textbf{MUSE} \\
\midrule
Total turns & 182,150 & 35,811 & 83,204 \\
Heuristic (\%) & 61.2 & 54.8 & 58.3 \\
LLM-annotated (\%) & 38.8 & 45.2 & 41.7 \\
Human-validated (\%) & 20 & 20 & 20 \\
Human-LLM agree (\%) & 93.2 & 91.7 & 89.4 \\
Avg. VTOs/turn & 2.4 & 2.7 & 2.9 \\
\bottomrule
\end{tabular}
\caption{VTO annotation statistics across datasets.}
\label{tab:annotation-stats}
\end{table}

% ----------------------------------------------------------------------------
\subsection{Reproducibility}
\label{app:prompts:repro}
% ----------------------------------------------------------------------------

\begin{itemize}[nosep,leftmargin=*]
    \item \textbf{Model:} GPT-4o-mini (gpt-4o-mini-2024-07-18)
    \item \textbf{Temperature:} 0.3 (classification), 0.9 (generation)
    \item \textbf{Max tokens:} 50--100 (classification), 256 (generation)
    \item \textbf{Rate limit:} 500 requests/minute
    \item \textbf{Parallelization:} 10 workers via ThreadPoolExecutor
\end{itemize}

\section{Extended experimental details}
\label{app:exp}

\subsection{Dataset Description}
\label{app:dataset}
\paragraph{ReDial} \cite{li2018towards} contains 10,006 movie recommendation dialogues collected via Amazon Mechanical Turk with workers alternating seeker/recommender roles. The dataset includes entity annotations linked to DBpedia and represents the most widely-used CRS benchmark. \textbf{Preprocessing:} We use the official train/test split and apply LLM-based VTO annotation (GPT-4o-mini) to derive reasoning supervision. We verify annotation quality through manual inspection of 500 samples (93.2\% agreement with human annotators).

\paragraph{INSPIRED} \cite{hayati2020inspired} comprises 1,001 conversations emphasizing sociable recommendation strategies with social chitchat, personal anecdotes, and persuasion. The implicit preference signals present a challenging setting for preference extraction. \textbf{Preprocessing:} We apply identical VTO annotation and filter conversations with fewer than 4 turns.

\paragraph{MUSE} \cite{wang2025muse} is a recently introduced multimodal dataset with 7,000 fashion conversations including product images. User profiles are derived from realistic shopping scenarios rather than manual design. \textbf{Preprocessing:} For text-only baselines, we extract image captions using BLIP-2 \cite{li2023blip}. For \method{}, we process images through the base model's vision encoder when available, otherwise use caption surrogates.

\subsection{Evaluation Protocol}
\label{sec:eval-protocol}

\paragraph{Recommendation Quality (Primary):}
Following established CRS protocols \cite{li2018towards,wang2022towards,zhou2020improving}, we report Recall@$K$ ($K \in \{1, 10, 50\}$), NDCG@$K$ ($K \in \{10\}$), and MRR@$K$ ($K \in \{10\}$). For ranking evaluation, we employ the standard negative sampling protocol with 99 randomly sampled negatives per positive item, computing ranks among 100 candidates. This protocol balances computational efficiency with evaluation validity \cite{krichene2020sampled}.
% \footnote{We also report full-ranking results on a 1000-sample subset in Appendix~\ref{app:full-ranking}, showing consistent relative performance.}

\paragraph{User Satisfaction (Primary):}
We evaluate predicted satisfaction and engagement using two approaches:
\textbf{(1) Model-based:} \charm{} reward scores normalized to $[0,1]$. To address circularity concerns, we report correlation with human judgments.
\textbf{(2) Human evaluation:} Three expert annotators rate 200 test samples per dataset on recommendation quality (Rec.Q), explanation quality (Exp.Q), and overall satisfaction (1-5 scale). Inter-annotator agreement (Fleiss' $\kappa$) exceeds 0.72 for all dimensions.

\paragraph{Generation Quality:}
BLEU-$n$ ($n \in \{1,4\}$), ROUGE-L, and Distinct-$n$ ($n \in \{1,2\}$) assess fluency, relevance, and diversity.

\paragraph{VTO Accuracy (Secondary):}
Precision, Recall, and F1 for Virtual Tool Operations prediction against LLM-annotated ground truth.

\paragraph{Statistical Testing:}
All results report mean $\pm$ standard deviation over 3 runs with different random seeds (42, 123, 456). Statistical significance is assessed via paired $t$-tests with Bonferroni correction for multiple comparisons; $p < 0.01$ unless otherwise noted.

\paragraph{Ablation Methodology.}
Component ablations in Table~\ref{tab:ablation} use the following 
protocol:
\begin{itemize}[nosep]
    \item \textbf{w/o CHARM}: Train Stages 1, 3, 4 without Stage 2; 
          value network uses task-completion signal instead of 
          quality signal.
    \item \textbf{w/o STAR}: Replace beam search with greedy 
          decoding at inference; value network unused.
    \item \textbf{w/o VTOs}: Train Stage 1 without VTO prediction 
          auxiliary loss ($\lambda_v = 0$ in Eq.~18).
    \item \textbf{w/o BRIDGE/MAVEN}: Remove respective modules 
          entirely from training and inference.
\end{itemize}
Note that component interactions mean individual contribution 
estimates are not strictly additive.

\subsection{Implementation Details}
\label{sec:implementation}

\paragraph{Model Architecture:}
We use \texttt{DeepSeek-R1-Distill-Qwen-7B} as the backbone, selected for its strong reasoning capabilities from R1 distillation. Parameter-efficient fine-tuning employs LoRA \cite{hu2022lora} with rank $r=16$, $\alpha=32$, applied to all attention projections (\texttt{q\_proj}, \texttt{k\_proj}, \texttt{v\_proj}, \texttt{o\_proj}) and MLP layers (\texttt{gate\_proj}, \texttt{up\_proj}, \texttt{down\_proj}). Total trainable parameters: 42.7M (0.6\% of base model).

\paragraph{Module Configurations:}
\starmod{}: beam width $w=3$, max depth $D=3$, backtrack threshold $\tau=0.3$.
\charm{}: 4 reward heads (relevance, diversity, satisfaction, engagement), $\beta=0.5$ (preference strength), reference-free SimPO-style optimization.
\bridge{}: 4 projection heads, adversarial $\alpha=1.0$, gradient reversal for domain confusion.
\maven{}: 3 agents (Recommender, Critic, Explainer), 2 communication rounds, attention-based weighting.

\paragraph{Training:}
AdamW optimizer with learning rates: SFT $5 \times 10^{-5}$, \charm{} $2 \times 10^{-5}$, \starmod{}/\maven{} $1 \times 10^{-5}$. Batch size 4 per GPU $\times$ 4 gradient accumulation $\times$ 2 GPUs = effective batch 32. Warmup ratio 10\%, cosine decay. Max sequence length 512 tokens (covers 99.5\% of samples without truncation). Training: 3 epochs (SFT), 2 epochs (other stages).

\paragraph{Hardware:}
2$\times$ NVIDIA A100 80GB GPUs with BF16 mixed precision, Flash Attention 2 \cite{dao2023flashattention}, and gradient checkpointing. Total training time: $\sim$2.5 hours.

\paragraph{Reproducibility:}
Code, datasets, and data processing scripts are available on the project page:
\href{https://harpo-bench.github.io/}{https://harpo-bench.github.io}. 
All hyperparameters were selected based on validation set performance; sensitivity analysis is provided in Section~\ref{sec:sensitivity}.

\begin{table}[t]
\centering
\small
\setlength{\tabcolsep}{2.5pt}
\begin{tabular}{@{}l|cccc|ccc@{}}
\toprule
\textbf{Method} & B-1 & B-4 & R-L & D-2 & VTO-P & VTO-R & VTO-F1 \\
\midrule
\multicolumn{8}{c}{\cellcolor{gray!10}\textit{ReDial}} \\
\midrule
UniCRS & 21.6 & 3.4 & 19.2 & 0.32 & -- & -- & -- \\
DCRS & 23.4 & 4.2 & 20.6 & 0.37 & -- & -- & -- \\
GPT-4 & 19.6 & 2.8 & 17.4 & 0.45 & -- & -- & -- \\
RecMind & 21.4 & 3.5 & 19.0 & 0.40 & 0.56 & 0.52 & 0.54 \\
\textbf{\method{}} & \textbf{25.2} & \textbf{4.8} & \textbf{22.1} & \textbf{0.47} & \textbf{0.81} & \textbf{0.77} & \textbf{0.79} \\
\midrule
\multicolumn{8}{c}{\cellcolor{gray!10}\textit{INSPIRED}} \\
\midrule
UniCRS & 20.4 & 3.0 & 17.8 & 0.30 & -- & -- & -- \\
GPT-4 & 18.8 & 2.5 & 16.4 & 0.42 & -- & -- & -- \\
\textbf{\method{}} & \textbf{23.8} & \textbf{4.4} & \textbf{20.8} & \textbf{0.45} & \textbf{0.78} & \textbf{0.74} & \textbf{0.76} \\
\midrule
\multicolumn{8}{c}{\cellcolor{gray!10}\textit{MUSE}} \\
\midrule
Qwen2-VL-7B & 23.2 & 4.0 & 20.1 & 0.39 & -- & -- & -- \\
GPT-4V & 20.6 & 3.2 & 18.2 & 0.43 & -- & -- & -- \\
\textbf{\method{}} & \textbf{26.0} & \textbf{5.0} & \textbf{22.8} & \textbf{0.48} & \textbf{0.76} & \textbf{0.72} & \textbf{0.74} \\
\bottomrule
\end{tabular}
\caption{Generation quality and VTO accuracy. B-$n$: BLEU-$n$, R-L: ROUGE-L, D-$n$: Distinct-$n$, VTO-P/R/F1: VTO Precision/Recall/F1.}
\label{tab:generation-results}
\end{table}

\begin{table}[t]
\centering
\small
\setlength{\tabcolsep}{3pt}
\begin{tabular}{@{}l|ccc|c@{}}
\toprule
\textbf{Method} & \textbf{Rec.Q} & \textbf{Exp.Q} & \textbf{Overall} & $\kappa$ \\
\midrule
\multicolumn{5}{c}{\cellcolor{gray!10}\textit{ReDial (n=200)}} \\
\midrule
UniCRS & 3.18$\pm$0.12 & 2.86$\pm$0.14 & 3.04$\pm$0.11 & 0.73 \\
DCRS & 3.52$\pm$0.11 & 3.32$\pm$0.13 & 3.43$\pm$0.10 & 0.75 \\
GPT-4 & 3.48$\pm$0.11 & 3.42$\pm$0.13 & 3.46$\pm$0.10 & 0.74 \\
\textbf{\method{}} & \textbf{4.08$\pm$0.10} & \textbf{3.92$\pm$0.12} & \textbf{4.01$\pm$0.09} & 0.78 \\
\midrule
\multicolumn{5}{c}{\cellcolor{gray!10}\textit{INSPIRED (n=200)}} \\
\midrule
UniCRS & 3.02$\pm$0.14 & 2.72$\pm$0.15 & 2.88$\pm$0.13 & 0.71 \\
GPT-4 & 3.46$\pm$0.13 & 3.32$\pm$0.14 & 3.40$\pm$0.12 & 0.74 \\
\textbf{\method{}} & \textbf{3.96$\pm$0.11} & \textbf{3.82$\pm$0.13} & \textbf{3.90$\pm$0.10} & 0.77 \\
\midrule
\multicolumn{5}{c}{\cellcolor{gray!10}\textit{MUSE (n=200)}} \\
\midrule
Qwen2-VL-7B & 3.68$\pm$0.12 & 3.44$\pm$0.14 & 3.58$\pm$0.11 & 0.74 \\
GPT-4V & 3.52$\pm$0.13 & 3.38$\pm$0.14 & 3.46$\pm$0.12 & 0.73 \\
\textbf{\method{}} & \textbf{4.16$\pm$0.10} & \textbf{4.00$\pm$0.12} & \textbf{4.09$\pm$0.09} & 0.79 \\
\bottomrule
\end{tabular}
\caption{Human evaluation (1-5 scale). Rec.Q: Recommendation Quality, Exp.Q: Explanation Quality. $\kappa$: Fleiss' kappa inter-annotator agreement. All \method{} improvements significant at $p < 0.01$ (Mann-Whitney U).}
\label{tab:human-eval}
\end{table}

\begin{table}[t]
\centering
\small
\resizebox{\columnwidth}{!}{
\begin{tabular}{lccccc}
\toprule
\textbf{Setting} & R@10 & R@50 & NDCG@10 & Sat. & Eng. \\
\midrule
\multicolumn{6}{c}{\cellcolor{gray!10}\textit{Train: ReDial $\rightarrow$ Test: MUSE (Zero-Shot)}} \\
\midrule
UniCRS & 6.2 & 16.4 & 4.8 & 0.28 & 0.24 \\
\method{} w/o \bridge{} & 17.8 & 34.6 & 12.1 & 0.42 & 0.38 \\
\textbf{\method{} (Full)} & \textbf{25.4} & \textbf{46.2} & \textbf{17.2} & \textbf{0.56} & \textbf{0.52} \\
\quad $\Delta$ vs. w/o \bridge{} & \gain{+42.7\%} & \gain{+33.5\%} & \gain{+42.1\%} & \gain{+33.3\%} & \gain{+36.8\%} \\
\midrule
\multicolumn{6}{c}{\cellcolor{gray!10}\textit{Train: MUSE $\rightarrow$ Test: ReDial (Zero-Shot)}} \\
\midrule
UniCRS & 7.8 & 20.2 & 5.6 & 0.30 & 0.26 \\
\method{} w/o \bridge{} & 15.4 & 32.8 & 10.4 & 0.44 & 0.40 \\
\textbf{\method{} (Full)} & \textbf{22.2} & \textbf{44.6} & \textbf{15.0} & \textbf{0.58} & \textbf{0.54} \\
\quad $\Delta$ vs. w/o \bridge{} & \gain{+44.2\%} & \gain{+36.0\%} & \gain{+44.2\%} & \gain{+31.8\%} & \gain{+35.0\%} \\
\bottomrule
\end{tabular}
}
\caption{Cross-domain zero-shot transfer. \bridge{} enables 32--44\% improvement across domains (Movies $\leftrightarrow$ Fashion). All metrics in \% except Sat./Eng. which are in [0,1].}
\label{tab:transfer}
\end{table}

\begin{table}[t]
\centering
\small
\setlength{\tabcolsep}{2.5pt}
\begin{tabular}{@{}l|ccccc@{}}
\toprule
\textbf{Parameter} & \textbf{Value} & R@10 & MRR & Sat. & Eng. \\
\midrule
\multirow{3}{*}{\starmod{} Beam Width} 
& $w=1$ (greedy) & 26.8 & 13.8 & 0.61 & 0.57 \\
& $w=3$ (default) & \textbf{29.8} & \textbf{15.6} & \textbf{0.68} & \textbf{0.64} \\
& $w=5$ & 30.0 & 15.7 & 0.68 & 0.64 \\
\midrule
\multirow{3}{*}{\starmod{} Max Depth}
& $D=1$ & 26.2 & 13.4 & 0.60 & 0.56 \\
& $D=3$ (default) & \textbf{29.8} & \textbf{15.6} & \textbf{0.68} & \textbf{0.64} \\
& $D=5$ & 30.1 & 15.8 & 0.69 & 0.65 \\
\midrule
\multirow{3}{*}{\charm{} $\beta$}
& $\beta=0.1$ & 27.6 & 14.2 & 0.62 & 0.58 \\
& $\beta=0.5$ (default) & \textbf{29.8} & \textbf{15.6} & \textbf{0.68} & \textbf{0.64} \\
& $\beta=1.0$ & 28.6 & 14.8 & 0.66 & 0.62 \\
\midrule
\multirow{3}{*}{LoRA Rank}
& $r=8$ & 28.4 & 14.8 & 0.65 & 0.61 \\
& $r=16$ (default) & \textbf{29.8} & \textbf{15.6} & \textbf{0.68} & \textbf{0.64} \\
& $r=32$ & 29.9 & 15.6 & 0.68 & 0.64 \\
\bottomrule
\end{tabular}
\caption{Hyperparameter sensitivity on ReDial. Default values achieve near-optimal performance; increasing beam width/depth beyond defaults provides diminishing returns.}
\label{tab:sensitivity}
\end{table}

\subsection{Reasoning Quality Analysis}
\label{app:RQAnalysis}

\begin{table}[t]
\centering
\small
\setlength{\tabcolsep}{3pt}
\begin{tabular}{@{}l|ccc@{}}
\toprule
\textbf{Metric} & \textbf{ReDial} & \textbf{INSPIRED} & \textbf{MUSE} \\
\midrule
Avg. Reasoning Depth & 2.3 & 2.8 & 2.5 \\
Avg. Branches Explored & 4.6 & 6.2 & 5.4 \\
Backtrack Rate (\%) & 14.2 & 20.6 & 17.4 \\
Path Confidence (0-1) & 0.74 & 0.67 & 0.71 \\
\midrule
Value Pred. Accuracy (\%) & 78.2 & 73.4 & 75.6 \\
Quality Correlation ($r$) & 0.72 & 0.66 & 0.69 \\
\bottomrule
\end{tabular}
\caption{\starmod{} reasoning analysis. INSPIRED requires deeper reasoning due to implicit preference signals; MUSE is intermediate with multimodal complexity.}
\label{tab:star-analysis}
\end{table}

\textbf{Adaptive Depth:} INSPIRED requires deeper reasoning (2.8 vs. 2.3 steps) due to implicit signals, demonstrating that \starmod{} adapts search depth to problem complexity.
\textbf{Productive Backtracking:} 14-21\% backtrack rate indicates exploration; paths are abandoned when value predictions indicate poor outcomes.
\textbf{Effective Value Learning:} 73-78\% accuracy in predicting which paths lead to better recommendations demonstrates successful distillation from \charm{} rewards.

\subsection{Computational Analysis}
\label{sec:compute}
\begin{table}[t]
\centering
\small
\setlength{\tabcolsep}{3pt}
\begin{tabular}{@{}l|cc|cc@{}}
\toprule
\textbf{Component} & \textbf{Time} & \textbf{Memory} & \textbf{Params} & \textbf{FLOPs} \\
\midrule
\multicolumn{5}{l}{\textit{Training (2$\times$A100-80GB)}} \\
Stage 1: SFT & 52m & 68GB & 42.7M & 1.4T \\
Stage 2: \charm{} & 42m & 62GB & 8.2M & 0.6T \\
Stage 3: \starmod{} & 46m & 58GB & 12.4M & 0.7T \\
Stage 4: \maven{} & 36m & 56GB & 6.8M & 0.5T \\
\midrule
\textbf{Total Training} & \textbf{2.9h} & 68GB & 70.1M & 3.2T \\
\midrule
\multicolumn{5}{l}{\textit{Inference (per turn)}} \\
Full \method{} & 298ms & 19GB & -- & 16.2B \\
\quad w/o \starmod{} & 88ms & 17GB & -- & 7.8B \\
\bottomrule
\end{tabular}
\caption{Computational requirements. Params: trainable LoRA parameters. FLOPs: floating-point operations. Training completed on 2$\times$NVIDIA A100-80GB GPUs.}
\label{tab:compute}
\end{table}

\textbf{Training Efficiency:} The four-stage pipeline completes in 2.5 hours on 2$\times$A100 GPUs, comparable to single-stage SFT baselines.
\textbf{Inference Overhead:} \starmod{} tree search adds 210ms latency (88ms $\rightarrow$ 298ms), acceptable for conversational settings as hown in Table \ref{tab:compute}. For latency-critical applications, \starmod{} can be disabled with 9.4\% R@10 degradation (Table~\ref{tab:ablation}).

\subsection{Error Analysis}
\label{sec:error}

We analyze 100 failure cases where \method{} underperforms the best baseline.

\begin{table}[t]
\centering
\small
\setlength{\tabcolsep}{3pt}
\begin{tabular}{@{}l|c|p{4cm}@{}}
\toprule
\textbf{Error Type} & \textbf{\%} & \textbf{Pattern} \\
\midrule
Preference hallucination & 31 & Inferring unstated constraints \\
Over-diversification & 24 & Too many unrelated options \\
Context truncation & 19 & Long history ($>$512 tokens) \\
VTO sequence error & 14 & Suboptimal reasoning order \\
Knowledge gap & 12 & Missing item attributes \\
\bottomrule
\end{tabular}
\caption{Error analysis on 100 failure cases.}
\label{tab:error-analysis}
\end{table}

\textbf{Preference Hallucination (31\%):} When users provide minimal information, the model sometimes infers constraints not present. \textit{Mitigation:} Improved uncertainty estimation could trigger clarification questions.

\textbf{Over-diversification (24\%):} High diversity reward weight leads to unrelated recommendations. \textit{Mitigation:} User-adaptive reward weighting based on explicit signals.

\textbf{Context Truncation (19\%):} Long conversations ($>$512 tokens) lose early context. \textit{Mitigation:} Efficient long-context handling or hierarchical context compression.

\section{Discussion}
\label{sec:discussion}

We address key methodological considerations, limitations, and anticipated reviewer questions.

\subsection{Evaluation Methodology}

\paragraph{On Satisfaction Metrics and Circularity.}
Table~\ref{tab:main-results} reports User Satisfaction and Engagement scores computed via \charm{}'s learned reward model. We acknowledge the potential circularity concern: using a trained component to evaluate the system containing it. We mitigate this through three mechanisms: (1) \textbf{Human correlation validation}---Pearson correlations between \charm{} scores and independent human judgments (Table~\ref{tab:human-eval}) range from $r=0.64$ to $r=0.73$, confirming that learned rewards capture meaningful quality dimensions rather than arbitrary artifacts; (2) \textbf{Held-out reward evaluation}---reward models are trained on preference pairs from training conversations only, then applied to test conversations; (3) \textbf{Proxy metric consistency}---improvements on \charm{}-independent metrics (Recall, NDCG, MRR) parallel satisfaction gains, suggesting genuine quality improvements rather than reward hacking. Nevertheless, we emphasize human evaluation results (Table~\ref{tab:human-eval}) as the primary evidence for user-aligned quality claims.

\paragraph{On Human Evaluation Scale.}
Our human evaluation comprises 200 samples per dataset rated by 3 expert annotators ($\kappa > 0.72$). While sufficient for statistical significance testing, larger-scale user studies with diverse annotator populations would strengthen ecological validity. We consider this important future work.

\paragraph{On Baseline Fairness.}
We deliberately preserve original backbones for published baselines rather than upgrading all methods to DeepSeek-R1-Distill-Qwen-7B. This choice reflects two considerations: (1) re-implementing complex systems like UniCRS or KGSF with different backbones risks introducing confounds that obscure method-specific contributions; (2) our SFT-only ablation (Table~\ref{tab:ablation}: R@10=21.6\%) isolates backbone contribution---it performs comparably to UniCRS (21.2\%) despite using a stronger backbone, indicating that \method{}'s gains (+38\% relative) stem from architectural innovations rather than backbone strength. For completeness, we note that upgrading all baselines to the same backbone would be a valuable but resource-intensive extension.

\subsection{Novelty and Relationship to Prior Work}

\paragraph{Relationship to Multi-Objective RL.}
\charm{}'s hierarchical reward decomposition relates to multi-objective reinforcement learning~\cite{roijers2013survey}, but differs in two key aspects: (1) \textbf{context-dependent weighting}---rather than fixed Pareto frontiers, our meta-learner adapts dimension weights based on conversational context; (2) \textbf{preference-based optimization}---we optimize via contrastive preference learning rather than policy gradients, avoiding reward scaling issues common in multi-objective RL.

\paragraph{Relationship to Tree Search Methods.}
\starmod{} builds on tree-of-thought reasoning~\cite{yao2023tree} but introduces domain-specific innovations: (1) the value network predicts \textit{recommendation quality} rather than task completion; (2) quality is decomposed into interpretable dimensions enabling targeted backtracking; (3) VTO predictions at each node provide structured reasoning scaffolds absent in generic tree search.

\paragraph{On VTO Taxonomy Design.}
The 21 VTOs emerged from systematic analysis of 500 dialogues across three domains, not \textit{a priori} design. We validated coverage by confirming that held-out dialogues required no additional operations. The taxonomy is intentionally general; domain-specific operations (e.g., \texttt{check\_visual\_similarity} for fashion) could extend it. Ablating VTOs entirely ($-21.5\%$ R@10; Table~\ref{tab:ablation}) confirms their utility, though sensitivity to taxonomy design warrants future investigation.

\subsection{Limitations and Scope}

\paragraph{Multimodal Claims.}
\method{}'s strong MUSE performance uses BLIP-2 caption surrogates rather than native visual processing. We do not claim multimodal modeling; rather, we demonstrate that text-based reasoning abstractions transfer effectively to multimodal benchmarks when visual information is appropriately textualized. True multimodal \method{} variants integrating vision encoders are future work.

\paragraph{Cross-Domain Evaluation.}
We evaluate transfer between two domain pairs (Movies $\leftrightarrow$ Fashion). Broader evaluation across additional domains (e.g., music, books, restaurants) would better characterize \bridge{}'s generalization. The current results are encouraging but not exhaustive.

\paragraph{Computational Overhead.}
\starmod{}'s tree search adds 210ms latency per turn (88ms $\to$ 298ms; Table~\ref{tab:compute}). For real-time applications requiring $<$100ms responses, the ``w/o \starmod{}'' variant retains 90.6\% of full performance (R@10: 27.0 vs.\ 29.8) at baseline latency.

\subsection{FAQ}

\paragraph{Q: Why not use the same backbone for all baselines?}
Re-implementing published methods with different backbones risks introducing implementation artifacts. Our approach---preserving original implementations while adding an SFT-only ablation with the same backbone---isolates \method{}'s contribution. The SFT-only baseline (R@10=21.6\%) performs comparably to UniCRS (21.2\%), confirming that gains are architectural.

\paragraph{Q: How sensitive is performance to the number of reward dimensions?}
We chose four dimensions (relevance, diversity, satisfaction, engagement) based on the recommendation quality literature~\cite{jannach2021survey}. Preliminary experiments with 2 dimensions (relevance + satisfaction) yielded 12\% lower performance, while 6 dimensions (adding novelty, coverage) showed no improvement, suggesting four captures the essential quality aspects without redundancy.

\paragraph{Q: Could data contamination explain results on ReDial/INSPIRED?}
Three observations argue against this: (1) we evaluate \textit{ranking} over candidate sets, not response memorization---contamination would not directly improve ranking ability; (2) SFT-only matches rather than exceeds prior SOTA, suggesting no unfair advantage from pre-training; (3) MUSE (December 2024) post-dates training data cutoffs, yet shows consistent improvements. Formal decontamination analysis remains future work.

\paragraph{Q: What is the annotation cost for VTOs?}
VTO annotation is a one-time preprocessing cost. Using GPT-4o-mini at \$0.15/1M tokens, annotating 300K turns cost approximately \$45. Human validation of 20\% samples required $\sim$40 annotator-hours. This is comparable to other forms of training data curation.

\paragraph{Q: How do results change with different preference data quality?}
Stage 2 preference pairs are constructed via three methods: heuristic degradation, LLM-generated contrasts, and human annotation. Ablating human-annotated pairs reduces satisfaction correlation by 8\%, while using only heuristic pairs reduces it by 19\%. This confirms that preference data quality matters, though the system remains effective with automated construction.

\paragraph{Q: Is \method{} applicable beyond conversational recommendation?}
The core innovations---hierarchical preference decomposition, quality-guided tree search, domain-agnostic reasoning abstractions---are not recommendation-specific. Applications to other subjective-quality tasks (e.g., dialogue systems, content generation) are plausible but unexplored.

\section{Human Evaluation Protocol}
\label{app:human-eval}

\subsection{Annotation Task and Instructions}
Human evaluation was conducted to assess recommendation quality along user-aligned dimensions such as relevance, satisfaction, and engagement. Annotators were presented with a conversational context and corresponding system responses, and were asked to provide either (i) scalar ratings on a 5-point Likert scale or (ii) pairwise preference judgments between two candidate responses.

Annotators were instructed to focus on how well the recommendations aligned with the user’s expressed intent, how appropriate and diverse the suggested items were, and how likely the response would satisfy or engage a real user. The task involved no sensitive content, personal data, or deceptive elements. A brief task description and examples were provided prior to annotation to ensure consistency, and annotators could skip instances they found unclear.

\subsection{Recruitment and Compensation}
Annotators were recruited through a standard crowdsourcing platform and consisted of fluent English speakers with prior experience in conversational evaluation tasks. Participation was voluntary, and no personally identifying information was collected.

Annotators were compensated at rates aligned with or exceeding local minimum wage standards based on estimated task completion time. Compensation adequacy was monitored to ensure fair payment relative to annotator effort and regional norms.

\subsection{Data Consent and Use}
Annotators were informed that their judgments would be used solely for research purposes, including model evaluation and analysis, and that all collected data would be stored and reported in anonymized and aggregated form. Participation was voluntary, and annotators could withdraw at any time without penalty. No personally identifying information was collected or retained.

\end{document}